\documentclass{emulateapj}
\usepackage{graphicx}
\begin{document}

\def\etal{et al.\ \rm}
\def\ba{\begin{eqnarray}}
\def\ea{\end{eqnarray}}
\def\etal{et al.\ \rm}

\title{Dynamical evolution of thin dispersion-dominated 
planetesimal disks}

\author{Roman R. Rafikov\altaffilmark{1,2} \& 
Zachary S. Slepian\altaffilmark{1}}
\altaffiltext{1}{Department of Astrophysical Sciences, 
Princeton University, Ivy Lane, Princeton, NJ 08540; 
rrr@astro.princeton.edu}
\altaffiltext{2}{Sloan Fellow}

%%%%%%%%%%%%%%%%%%%%%%%%%%%%%%%%%%%%%%%%%%%%%%%%%%%%%%%%%%%

\begin{abstract}
We study the dynamics of a vertically thin, dispersion-dominated 
disk of planetesimals with eccentricities $\tilde e$ and inclinations 
$\tilde i$ (normalized in Hill units) 
satisfying $\tilde e\gg 1$, $\tilde i\ll \tilde e^{-2}\ll 1$. 
This situation may be 
typical for e.g. a population of protoplanetary cores in the 
end of the oligarchic 
phase of planet formation. In this regime of orbital parameters
planetesimal scattering has an anisotropic character and strongly differs 
from scattering in thick ($\tilde i\sim \tilde e$) disks. We derive analytical
expressions for the planetesimal scattering coefficients and compare 
them with numerical calculations. We find significant discrepancies 
in the inclination scattering coefficients obtained by the two 
approaches and ascribe this difference to the effects not accounted for 
in the analytical calculation: multiple scattering
events (temporary captures, which may be relevant for the production of 
distant planetary satellites outside the Hill sphere) and distant 
interaction of planetesimals prior to their close encounter. Our 
calculations show that the inclination of a thin, dispersion-dominated 
planetesimal disk grows exponentially on a very short time scale 
implying that (1) such disks must be very short-lived 
and (2) planetesimal accretion in this dynamical phase is insignificant.
Our results are also applicable to the dynamics of shear-dominated disks 
switching to the dispersion-dominated regime.
\end{abstract}

%%%%%%%%%%%%%%%%%%%%%%%%%%%%%%%%%%%%%%%%%%%%%%%%%%%%%%%%%%%
\section{Introduction.}  
\label{sect:intro}

Terrestrial planets are thought to be formed by 
agglomeration of a large number of primitive rocky or icy 
bodies known as planetesimals (Safronov 1972). While
the origin of planetesimals themselves is still a rather uncertain issue 
(Youdin 2008) the process of their collisional agglomeration 
has been extensively explored (Wetherill \& Stewart 1989, 1993; 
Kenyon \& Luu 1998; Kenyon \& Bromley 2004, 2009). 
Gravitationally induced 
bending of the trajectories of interacting bodies called gravitational 
focusing (Safronov 1972) is known to play a very important role in speeding 
up the agglomeration process. The degree to which the planetesimal collision 
rate is amplified by focusing depends sensitively on the velocity dispersion
of the planetesimals: the lower is the relative velocity between the 
interacting bodies the higher are the gravitational focusing and 
the collision cross-section. Thus, understanding the accretion history of 
planetesimals is impossible without understanding their dynamical 
evolution. 

Evolution of planetesimal velocities is driven mainly by their mutual 
gravitational interaction. A convenient way to characterize the shape 
of planetesimal orbits and their interaction is via 
the so-called eccentricity and inclination vectors ${\bf e}$ and 
${\bf i}$ defined as (Ida 1990)
\ba
&& {\bf e}=(e_x,e_y)=(e\cos\tau,e\sin\tau),\nonumber\\
&& {\bf i}=(i_x,i_y)=(i\cos\omega,i\sin\omega),
\label{eq:vectors}
\ea
where $e$, $i$, $\tau$ and $\omega$ are, respectively, the eccentricity, 
inclination, and 
horizontal and vertical phases of the planetesimal. Scattering
of two low-mass planetesimals depends only on their {\it relative} 
eccentricity and inclination vectors ${\bf e}_r={\bf e}_1-{\bf e}_2$ 
and ${\bf i}_r={\bf i}_1-{\bf i}_2$ (the so-called Hill approximation, see
H\'enon \& Petit 1986).

There are two important asymptotic regimes of planetesimal interaction: 
{\it shear-dominated} and {\it dispersion-dominated}. The former is 
realized when the random component of planetesimal velocity, determined 
by its eccentricity and inclination, is small compared to 
the Hill velocity $v_H=\Omega R_H$.
Here $\Omega$ is the local angular frequency in the disk,  
$R_H\equiv a(\mu_1+\mu_2)^{1/3}$ is the Hill radius, determined by
the distance $a$ to the central object and the masses of the interacting 
bodies $m_1$ and $m_2$ relative to the central mass $M_\star$:
$\mu_i\equiv m_i/M_\star,~i=1,2$. Introducing scaled relative 
eccentricity\footnote{In this paper all quantities with a tilde are assumed
to be scaled by the Hill factor $(\mu_1+\mu_2)^{1/3}$.} 
$\tilde {\bf e}_r$ and inclination $\tilde {\bf i}_r$ 
vectors of the interacting bodies as\footnote{In this paper we use the  
Hill factor adopted by H\'enon \& Petit (1986), which differs by  
$3^{1/3}$ from the scaling used by some other authors.} 
$\tilde {\bf e}_r\equiv {\bf e}_r/(\mu_1+\mu_2)^{1/3}$ 
and $\tilde {\bf i}_r\equiv {\bf i}_r/(\mu_1+\mu_2)^{1/3}$, 
one can rewrite the condition for the shear-dominated regime as 
\ba
\tilde e_r^2+\tilde i_r^2\lesssim 1.
\label{eq:sd}
\ea
The relative speed of a pair of interacting bodies in this
regime is set mainly by the Keplerian shear. 

Dispersion-dominated regime of planetesimal interaction is realized when
\ba
\tilde e_r^2+\tilde i_r^2\gtrsim 1.
\label{eq:dd}
\ea
In this case the relative velocity of the planetesimals is determined mainly by
their random epicyclic motion while Keplerian shear plays only a minor role.
This makes possible analytical treatment of planetesimal dynamics 
(Ida 1990; Ida \& Makino 1992; Tanaka \& Ida 1996,1997; 
Stewart \& Ida 2000), which until now has been concentrated on
the case when $\tilde i \sim \tilde e$, so that the random 
velocity distribution
of planetesimals is roughly isotropic. This assumption is very natural 
in advanced stages of dynamical evolution of the dispersion-dominated 
planetesimal population but it may fail in more general situations. 

It is thought (Kokubo \& Ida 1998; Rafikov 2004; 
Goldreich \etal 2004) that at the very end of the oligarchic stage 
of planetary growth in the inner parts of the Solar System, just before 
the transition to a chaotic final stage of planetary assembly (sometimes 
called the stage of giant impacts), planetesimal coagulation 
produced a number (several hundred) of protoplanetary cores 
with masses comparable to the mass of the Moon or Mars, i.e. 
$\sim (0.01-0.1)$ M$_\oplus$. These cores comprised a significant 
fraction of all the refractory mass of the disk and were well-separated 
in semi-major axis (typically by several Hill radii).

The orbits of these cores are initially not expected to cross
because their eccentricities are  
very small as a result of efficient dynamical friction 
exerted on them by the residual population of small planetesimals.
However, with time the population of small planetesimals gets eroded
by collisional grinding and accretion by cores
and the strength of dynamical friction goes down. Distant mutual 
gravitational perturbations between nearby cores then gradually 
increase their velocity dispersion, eventually allowing their orbits 
to cross, which leads to collisions between embryos and their growth into 
bigger bodies. 

Since initially the inclinations of the cores were almost zero distant
perturbations cannot efficiently excite vertical motion of cores. Moreover, 
even though dynamical friction from the remaining planetesimals is no 
longer efficient in curbing the eccentricity growth of the cores it may still
be strong enough to continue damping their inclinations. As a result, the 
protoplanetary cores are expected to reside in a very thin disk with 
$\tilde i\ll 1$ all the way until the point when their orbits  
start to 
cross. When this happens one finds that $\tilde i\ll 1\lesssim \tilde e$,
so that the condition $\tilde i\sim\tilde e$ usually assumed in studies of 
planetesimal dynamical evolution is strongly violated.

Thus, a vertically thin, dispersion-dominated 
planetesimal disk can naturally arise in some circumstances. 
Since the collision rate of planetesimals is 
a sensitive function of their inclination -- the smaller is inclination the 
higher is the collision probability and the faster is protoplanetary 
growth -- it is important to know how much time the population
of cores spends in the thin disk configuration after their orbits 
become crossing. If this time is sufficiently long then core masses 
could grow significantly by collisions even during the transient period 
when their inclinations have not yet increased. This possibility 
potentially may act to speed up the final assembly of terrestrial 
planets.

A similar situation arises when one considers the transition of the 
shear-dominated planetesimal population into the dispersion-dominated regime.
Ida \& Makino (1992) showed that a planetesimal population 
starting in the shear-dominated regime typically undergoes a phase in 
its dynamical evolution when $\tilde i\ll 1\lesssim  \tilde e$ 
(see their Fig. 6 for 
illustration). This phase does not persist for very long but while 
it lasts the dynamics of the planetesimals are significantly different from 
the usually assumed case of $\tilde i\sim \tilde e$. 

These considerations give us a motivation to explore the
dynamical regime $\tilde i\ll 1\lesssim e$ in this work.
The paper is organized as follows: in \S \ref{sect:velev} we describe 
the equations governing the velocity evolution of the planetesimals while in 
\S \ref{sect:scat_coef} we analytically compute the scattering coefficients 
entering these equations in the case of $\tilde i\ll 1\lesssim \tilde e$,
and in \S \ref{sect:numerics} we compare our results with numerical 
calculations. In \S \ref{subsect:velev} we use our results to examine 
the velocity evolution of a population of protoplanetary cores with 
crossing orbits. In \S \ref{sect:disc} we provide comparison with 
other studies and discuss additional applications of our results.

%%%%%%%%%%%%%%%%%%%%%%%%%%%%%%%%%%%%%%%%%%%%%%%%%%%%%%%%%%%
\section{Velocity evolution.}
\label{sect:velev}

To understand the velocity evolution of planetesimals we consider two 
populations of planetesimals with masses $m_1$ and $m_2$; 
populations of different mass 
contribute linearly to velocity evolution so it is sufficient 
to consider just two masses. We assume that for every planetesimal 
type $e_k,i_k\ll 1$ and $m_k\ll M_\star$, $k=1,2$, providing 
justification for using the Hill approximation. 

In this local approximation the Keplerian orbit of a $k$-th 
planetesimal type is described by 
the following equations:
\ba
&& x_k=h_k-e_k\cos(t-\tau_k),
\label{eq:x}\\
&& y_k=\lambda_k-\frac{3}{2}h_k t+2e\sin(t-\tau_k),
\label{eq:y}\\
&& z_k=i_k\sin(t-\omega_k),
\label{eq:z}
\ea
where $x$, $y$, and $z$ are Cartesian coordinates in the local radial, 
azimuthal, and vertical directions centered at some reference 
stellocentric distance, $h$ is the planetesimal semi-major 
axis separation from the origin of this coordinate system, and 
$\lambda$ is a constant related to the origin of time $t$ 
(measured in units of $\Omega^{-1}$). 

The relative motion of two non-interacting planetesimals in Hill units 
($\tilde {\bf r}_r=({\bf r}_1-{\bf r}_2)/a(\mu_1+\mu_2)^{1/3}$) is given by 
equations
\ba
&& \tilde x_r=\tilde h_r-\tilde e_r\cos(t-\tau_r),
\label{eq:xH}\\
&& \tilde y_r=\tilde \lambda_r-\frac{3}{2}\tilde h_r t+2\tilde 
e_r\sin(t-\tau_r),
\label{eq:yH}\\
&& \tilde z_r=\tilde i_r\sin(t-\omega_r), 
\label{eq:zH}
\ea
where $\tilde e_r$, $\tilde i_r$ are the relative eccentricity and inclination
of the planetesimals, $\tilde h\equiv (a_1-a_2)/a(\mu_1+\mu_2)^{1/3}$ is
the semimajor axes separation normalized in Hill units, and 
$\tilde \lambda_r=(\lambda_1-\lambda_2)/a(\mu_1+\mu_2)^{1/3}$.
In the following we will drop the subscript ``r'' from all variables 
characterizing relative motion of planetesimals where it will not lead 
to confusion. 

Because of the mutual gravitational attraction relative orbital elements 
appearing in equations (\ref{eq:xH})-(\ref{eq:zH})
do not remain constant but change according to the following set of 
equations (Hasegawa \& Nakazawa 1990; Tanaka \& Ida 1996):
\ba
&& \frac{d \tilde h}{d t}=-2\frac{\partial \phi}{\partial \tilde y},
\label{eq:hdot}\\
&& \frac{d \tilde \lambda}{d t}=2\frac{\partial \phi}
{\partial \tilde x}-3 t\frac{\partial \phi}{\partial \tilde y},
\label{eq:lamdot}\\
&& \frac{d \tilde e_{x}}{d t}=-\sin t\frac{\partial \phi}
{\partial \tilde x}-2\cos t\frac{\partial \phi}{\partial \tilde y},
\label{eq:exdot}\\
&& \frac{d \tilde e_{y}}{d t}=\cos t\frac{\partial \phi}
{\partial \tilde x}-2\sin t\frac{\partial \phi}{\partial \tilde y},
\label{eq:eydot}\\
&& \frac{d \tilde i_{x}}{d t}=-\cos t\frac{\partial \phi}
{\partial \tilde z},
\label{eq:ixdot}\\
&& \frac{d \tilde i_{y}}{d t}=-\sin t\frac{\partial \phi}
{\partial \tilde z},
\label{eq:iydot}
\ea
where
\ba
\phi=-(\tilde x^2+\tilde y^2+\tilde z^2)^{-1/2}.
\label{eq:potential}
\ea
is the interaction potential.

In this work we assume for simplicity that ${\bf e}$ 
and ${\bf i}$ of a $k$-th planetesimal population have 
Gaussian distribution:
\begin{equation}
\psi({\bf e}_k,{\bf i}_k)d{\bf e}_k d{\bf i}_k=\frac{d{\bf e}_k 
~d{\bf i}_k}{4\pi^2 \sigma_{e,k}^2
\sigma_{i,k}^2}\exp\left[-\frac{{\bf e}_k^2}{2\sigma_{e,k}^2}
-\frac{{\bf i}_k^2}{2\sigma_{i,k}^2}\right],
\label{eq:Gauss}
\end{equation}
where $\sigma_{e,k}$ and $\sigma_{i,k}$ are the dispersions of 
eccentricity and inclination of the $k$-th population. Ida \& Makino (1992)
have found that a Gaussian distribution accurately describes the 
distribution of ${\bf e}$ and ${\bf i}$ found in direct N-body 
three-dimensional (3D) simulations of a large number of 
planetesimals gravitationally interacting in the dispersion-dominated 
regime. At the same time in their dispersion-dominated simulations 
of 2D disks Ida \& Makino found more high-energy particles than a 
Gaussian distribution would predict. Despite this we still use distribution 
(\ref{eq:Gauss}) to represent velocities of planetesimals in thin disks as
this is not going to strongly affect the velocity evolution but allows 
significant simplification.

Our goal is to find how $\sigma_{e,1}$ 
and $\sigma_{i,1}$ of a population with mass $m_1$ varies in time
as a result of gravitational interactions with planetesimals of mass $m_2$
which have eccentricity and inclination dispersions $\sigma_{e,2}$ 
and $\sigma_{i,2}$ 
(for now we neglect other factors that may affect planetesimal 
velocities such as gas drag, inelastic collisions, and so on). 
General evolution equations for the case of distribution 
(\ref{eq:Gauss}) have been previously 
derived by a number of authors (Hornung \etal 1985; Ida 1990; 
Wetherill \& Stewart 1993; Stewart \& Ida 2000). Here we adopt 
a specific expression from Rafikov (2003):
\begin{eqnarray}
&& \frac{\partial \sigma_{e,1}^2}{\partial t}\Big|_2=
\frac{3}{4}\Omega N_2 a^2
(\mu_1+\mu_2)^{4/3}
\nonumber\\
&& \times\left[
\left(\frac{\mu_2}{\mu_1+\mu_2}\right)^2 H_1
+2\frac{\mu_2}{\mu_1+\mu_2}\frac{\sigma_{e,1}^2}{\sigma_{e,1}^2+ 
\sigma_{e,2}^2}H_2\right],
\label{eq:homog_heating}
\end{eqnarray}
where $N_2$ is the surface number density of bodies with mass 
$m_2$. 

Dimensionless stirring coefficients $H_{1,2}$ appearing in equation 
(\ref{eq:homog_heating}) are defined as
\begin{eqnarray}
&& H_{k}=
\int d\tilde e d\tilde i \tilde
\psi_r(\tilde e,\tilde i)\hat H_{k}
(\tilde e,\tilde i),~~~k=1,2,
\label{eq:stirring_coeffs}\\
&& \hat H_{1}\equiv
\int\limits_{-\infty}^{\infty}d\tilde h 
|\tilde h|\langle\left(\Delta {\bf \tilde e}\right)^2
\rangle_{\tau,\omega},
\label{eq:stirring_coeffs1}\\
&&  \hat H_{2}\equiv
\int\limits_{-\infty}^{\infty}d\tilde h |\tilde h|\langle
({\bf \tilde e}\cdot\Delta {\bf \tilde e})\rangle_{\tau,\omega}.
\label{eq:stirring_coeffs2}
\end{eqnarray}
Here $\Delta {\bf \tilde e}$ is the change of 
${\bf \tilde e}$ in the course of scattering, 
and $\langle ... \rangle_{\tau,\omega}\equiv (4\pi^2)^{-1}
\int\int d\tau d\omega$ implies averaging over the {\it relative} 
orbital phases characterizing vectors ${\bf \tilde e}$ and  
${\bf \tilde i}$. Function 
$\tilde \psi_r(\tilde e_r,\tilde i_r)$ is the distribution function of 
{\it relative} $\tilde e$, $\tilde i$ and can be shown (Stewart \& Ida 
2000; Rafikov 2003) to be given by 
\begin{equation}
\psi_r(\tilde e,\tilde i)d\tilde e d\tilde i
=\frac{\tilde e d\tilde e 
~\tilde i d\tilde i}{\tilde \sigma_{e}^2
\tilde \sigma_{i}^2}\exp\left[-\frac{\tilde e^2}{2\tilde \sigma_{e}^2}
-\frac{\tilde i^2}{2\tilde\sigma_{i}^2}\right],
\label{eq:Rayleigh}
\end{equation}
where $\tilde\sigma_{e}^2=(\sigma_{e,1}^2+
\sigma_{e,2}^2)/(\mu_1+\mu_2)^{2/3}$ and 
$\tilde\sigma_{i}^2=(\sigma_{i,1}^2+
\sigma_{i,2}^2)/(\mu_1+\mu_2)^{2/3}$ are the dispersions of 
{\it relative} eccentricity and inclination. 
Thus, functions $\hat H_1$ and $\hat H_2$ 
represent scattering coefficients for a planetesimal population 
with a single value of relative eccentricity $e$ and 
inclination $i$, while $H_1$ and $H_2$ are these coefficients 
averaged over the distribution (\ref{eq:Rayleigh}) of $e$ and $i$.

The term inside the brackets in equation (\ref{eq:homog_heating}) 
proportional to $H_1$ is
called {\it gravitational stirring} (Rafikov 2003) while the 
term proportional to $H_2$ is called {\it gravitational friction} 
and is different from the 
``dynamical friction'' used by other authors (Stewart \& Ida 2000; 
Ohtsuki \etal 2002).

Equation (\ref{eq:homog_heating}) also describes the self-stirring 
of population with mass $m_1$ if one changes the subscript ``2''
to ``1'' in its right hand side (which makes expression in brackets equal
to $(H_1+2H_2)/4$). For a continuous distribution of planetesimal masses 
equation (\ref{eq:homog_heating}) should be generalized by integrating 
the right hand side over the mass spectrum. 

Equations analogous to (\ref{eq:homog_heating}) can be written 
for the inclination
evolution by replacing ``e'' by ``i'' everywhere in 
equations (\ref{eq:homog_heating})-(\ref{eq:stirring_coeffs2})
and using scattering coefficients $\hat K_{1,2}$ and $K_{1,2}$ 
defined analogously to expressions
(\ref{eq:stirring_coeffs})-(\ref{eq:stirring_coeffs2}) instead 
of $\hat H_{1,2}$ and $H_{1,2}$).

%%%%%%%%%%%%%%%%%%%%%%%%%%%%%%%%%%%%%%%%%%%%%%%%%%%%%%%%%%%
\section{Scattering coefficients.}
\label{sect:scat_coef}

System (\ref{eq:homog_heating})-(\ref{eq:stirring_coeffs2}) is a rather 
general set of equations derived for a Gaussian distribution of orbital 
elements (\ref{eq:Gauss}). It allows one to determine how $\sigma_{e, k}$ 
and  $\sigma_{i, k}$ ($k=1,2$) evolve in time once 
coefficients $\tilde H_{1,2}$ and $\tilde K_{1,2}$ are known 
as functions of $\tilde\sigma_{e}$ and $\tilde\sigma_{i}$.
These coefficients have been previously calculated by a number 
of authors in the dispersion-dominated case under the assumption 
that $\tilde i\sim \tilde e$. However, as we demonstrate shortly, these
calculations become invalid once $\tilde i$ gets sufficiently small 
while $\tilde e\gg 1$. Thus our next step is to rederive scattering 
coefficients in the case of $\tilde i\ll \tilde e$ from the first 
principles. 

In doing so we will adopt an approach previously developed 
by Ida \etal (1993) for the dispersion-dominated regime. According 
to this method (1) the approach trajectory of one 
planetesimal far from another is represented as a 
straight line and (2) the effect of perturbations from the central 
mass on the gravitational scattering of the two planetesimals  
is neglected. There is also an implicit assumption that (3) the scattering
coefficients are dominated (or at least strongly contributed to) by 
those planetesimals whose trajectories pass very close to 
the perturber. If the latter assumption is fulfilled then the other 
two are quite natural.
Indeed, the first approximation works well because most of the 
perturbation to the orbit of the passing particle occurs near the point 
of closest approach of the interacting bodies. In this region 
the curvature of the planetesimal trajectory caused by epicyclic 
motion can be neglected, justifying the straight-line simplification. 
The second approximation works because the most significant contribution to 
scattering is due to trajectories passing very close to the 
perturber, within its
Hill radius, where the influence of the third, central body can be 
disregarded.

When $\tilde e\sim\tilde i\gg 1$ the flux of approaching planetesimals 
in the vicinity of any given perturber is roughly uniform 
on scales $\sim R_H$, and every decade in the initial impact 
parameter of interacting bodies contributes 
roughly equally to the scattering coefficients (Binney \& 
Tremaine 1987). This gives rise to appearance of the so-called 
Coulomb logarithm, $\ln\Lambda$, in expressions for the scattering 
coefficients. The argument $\Lambda$
is roughly the ratio of the maximum impact parameter 
$l_{\max}\sim \tilde i R_H$, beyond which the density of 
approaching planetesimals is non-uniform, 
to the impact parameter  
\ba
l_{min}\sim \frac{R_H}{\tilde e^2+\tilde i^2}, 
\label{eq:l_min}
\ea
at which the trajectories of incoming planetesimals 
experience large-angle deflection.

Previous calculations of the scattering coefficients in the 
dispersion-dominated regime assumed that $l_{max}\gg l_{min}$ meaning 
that $\ln\Lambda\gtrsim 1$. In this case 
the scattering calculation for closely approaching orbits with impact 
parameters $\sim l_{min}\ll R_H$ which can be done analytically 
approximates quite well (with logarithmic accuracy)
the full scattering coefficients so that the aforementioned assumption 
of the dominance of close encounters for scattering coefficient 
calculation is roughly fulfilled. Clearly, $l_{max}\gg l_{min}$ requires that 
$\tilde i\gg 1/\tilde e^2$, which is essentially a condition for the 
standard expressions for the dispersion-dominated scattering 
coefficients to be valid. 

\begin{figure}
\plotone{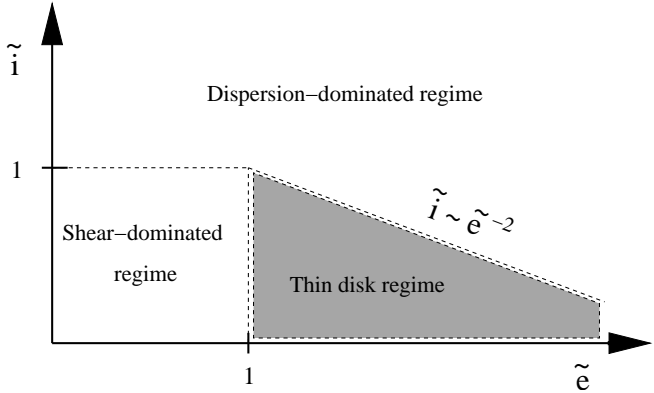}
\caption{
Schematic illustration of different regions in the $\tilde e$-$\tilde i$
phase space, showing the shear- and dispersion-dominated regions and
the thin-disk dynamical regime (shaded), which is also 
dispersion-dominated ($\tilde e\gtrsim 1$).
\label{fig:phase_space}}
\end{figure}

In this work we look at the opposite extreme, namely
a situation when 
\ba
\tilde i\lesssim \tilde i_{crit}\equiv \tilde e^{-2}\ll 1,~~~
\tilde e\gg 1,
\label{eq:condit}
\ea
(see Figure \ref{fig:phase_space} for illustration). 
When this condition is satisfied the assumption of a 
uniform distribution of approaching planetesimals around the scatterer 
does not hold even for $l\sim l_{min}$. Then a 
new calculation of scattering coefficients is needed. 

In Appendix A we present the details of such a calculation which 
makes the following set of assumptions: (1) 
the planetesimals move at high relative velocities which allows us
to use a two-body scattering approximation, (2) the planetesimal velocities 
change only during close approaches, which are possible only when 
$\tilde h<\tilde e$, (3) the gravitational interaction between planetesimals
at large separations is neglected, and (4) after changing as a result
of the encounter with the scatterer the 
planetesimal's orbital elements do not change further. These simplifications  
allow us to derive the following set of expressions for the 
integrands of the scattering coefficients, see equations 
(\ref{eq:stirring_coeffs1})-(\ref{eq:stirring_coeffs2}):
\begin{eqnarray}
&& \langle(\Delta\tilde {\bf e})^2\rangle_{\omega,\tau}=\frac{20}{3}
\frac{\tilde v}{|\tilde h|\sqrt{\tilde e^2-\tilde h^2}},
\label{eq:de2_omtau}\\
&& \langle\tilde {\bf e}\cdot\Delta\tilde {\bf e}\rangle_{\omega,\tau}
=-\frac{4}{3}\frac{\tilde e^2}{\tilde v|\tilde h|
\sqrt{\tilde e^2-\tilde h^2}},
\label{eq:ede_omtau}\\
&& \langle(\Delta\tilde {\bf i})^2\rangle_{\omega,\tau}=\frac{2}{3}
\frac{\tilde v^5 \tilde i^2}{|\tilde h|\sqrt{\tilde e^2-\tilde h^2}}.
\label{eq:di2_omtau}\\
&& \langle\tilde {\bf i}\cdot\Delta\tilde {\bf i}\rangle_{\omega,\tau}
=-\frac{2}{3}\frac{\tilde i^2}
{\tilde v|\tilde h|\sqrt{\tilde e^2-\tilde h^2}},
\label{eq:idi_omtau}
\end{eqnarray}
Note that these expressions do not contain a Coulomb logarithm and 
do not suffer from the ambiguity related to the choice 
of minimum and maximum values of impact parameter 
$l_{min}$ and $l_{max}$ typical for the 3D case. The physical reason 
for this lies in the fact that in the limit (\ref{eq:condit}) the scattering
coefficients are dominated by trajectories with impact parameters 
$l\sim l_{min}$, i.e. those leading to large-angle scattering. Thus,
integrals over $dl$ appearing in the calculation of the scattering 
coefficients are 
mostly contributed to by $l\sim l_{min}$ in the quasi-2D 
case\footnote{Note that
this property makes the assumptions adopted in our calculation quite robust.},
so that values of $l$ much larger and much smaller than $l_{min}$ affect 
the coefficients only weakly. This is very different from the 3D case, in 
which trajectories experiencing weak scattering provide 
significant contribution to the scattering coefficients. 

Integrating these equations over $|\tilde h|d\tilde h$ from 
$0$ to $\tilde e$ (limits within which a given planetesimal can experience 
a close encounter with the scatterer) and substituting into 
equations (\ref{eq:stirring_coeffs1})-(\ref{eq:stirring_coeffs2})
we arrive at the following expressions 
for the scattering coefficients corresponding to fixed $\tilde e$ and 
$\tilde i$ and averaged over the phase angles 
$\tau$ and $\omega$: 
\begin{eqnarray}
&& \hat H_1=
\frac{40}{3}~{\bf E}\left(\frac{\sqrt{3}}{2}\right)\tilde e_r=
16.147~\tilde e_r,
\label{eq:hatH_1}\\
&& \hat H_2=
-\frac{8}{3}~{\bf K}\left(\frac{\sqrt{3}}{2}\right)\tilde e_r=
-5.751~\tilde e_r ,~~~
\label{eq:hatH_2}\\
&& \hat K_1=
\frac{1}{9}\left[\frac{41}{5}{\bf E}\left(\frac{\sqrt{3}}{2}\right)
-{\bf K}\left(\frac{\sqrt{3}}{2}\right)\right]\tilde i_r^2 \tilde e_r^5
\nonumber \\
&& = 0.864~\tilde i_r^2 \tilde e_r^5,
\label{eq:hatK_1}\\
&& \hat K_2=
-\frac{4}{3}~{\bf K}\left(\frac{\sqrt{3}}{2}\right)
\frac{\tilde i_r^2}{\tilde e_r}=
-2.875~ \frac{\tilde i_r^2}{\tilde e_r}.
\label{eq:hatK_2}
\end{eqnarray} 
Finally, averaging coefficients (\ref{eq:hatH_1})-(\ref{eq:hatK_2}) 
over the Gaussian distribution (\ref{eq:Gauss}) one finds that
\begin{eqnarray}
H_1 &=& \frac{20\sqrt{2\pi}}{3}~{\bf E}\left(\frac{\sqrt{3}}{2}\right)\tilde 
\sigma_{e,r}\approx 20.237
~\tilde \sigma_{e,r},
\label{eq:H_1}\\
H_2 &=& -\frac{4\sqrt{2\pi}}{3}~{\bf K}\left(\frac{\sqrt{3}}{2}\right)\tilde 
\sigma_{e,r}\approx -7.207
~ \tilde \sigma_{e,r}\,
\label{eq:H_2}\\
K_1 &=& \frac{5\sqrt{2\pi}}{3}\left[\frac{41}{5}{\bf E}
\left(\frac{\sqrt{3}}{2}\right)
-{\bf K}\left(\frac{\sqrt{3}}{2}\right)\right] \tilde \sigma_{i,r}^2
\tilde \sigma_{e,r}^5\nonumber\\
&\approx & 
32.478
~\tilde \sigma_{i,r}^2\tilde \sigma_{e,r}^5,
\label{eq:K_1}\\
K_2 &=& -\frac{4\sqrt{2\pi}}{3}~{\bf K}\left(\frac{\sqrt{3}}{2}\right)
\frac{\tilde \sigma_{i,r}^2}{\tilde \sigma_{e,r}}\approx -7.207~
\frac{\tilde \sigma_{i,r}^2}{\tilde \sigma_{e,r}}
\label{eq:K_2}
\end{eqnarray}
These expressions represent the behavior of scattering 
coefficients in the limit (\ref{eq:condit}).

%%%%%%%%%%%%%%%%%%%%%%%%%%%%%%%%%%%%%%%%%%%%%%%%%%%%%%%%%%%
\section{Comparison with numerical results.}
\label{sect:numerics}

To check our analytical results we ran a series of numerical
calculations. The latter are the Monte-Carlo computations 
of integrals in equations 
(\ref{eq:stirring_coeffs})-(\ref{eq:stirring_coeffs2})
with $\Delta \tilde {\bf e}_r$, $\Delta \tilde {\bf i}_r$ obtained by 
direct integration of equations (\ref{eq:hdot})-(\ref{eq:iydot}). 
Equations for the evolution of orbital elements have been integrated 
using fourth-order Runge-Kutta method with adaptive step size control
(Press \etal 1992). Conservation of Jacobi constant has been routinely 
monitored and this integral of motion has been found to vary during 
the calculation by at most one part in $10^5$ for a very small 
number of orbits. In the majority of calculations the Jacobi constant 
has been conserved to relative accuracy of better than $10^{-10}$.

\begin{figure}
\plotone{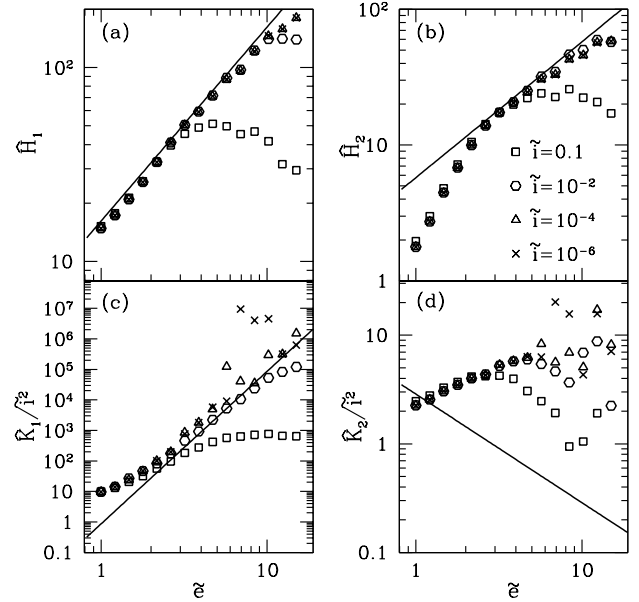}
\caption{
Results of numerical calculation of scattering coefficients (a) $\hat H_1$,
(b) $\hat H_2$, (c) $\hat K_1/\tilde i^2$, (d) $\hat K_2/\tilde i^2$, 
compared with analytical 
predictions (\ref{eq:hatH_1})-(\ref{eq:hatK_2}), represented by solid lines. 
Values of coefficients are shown as functions of $\tilde e$ for 
$\tilde i=10^{-1}, 10^{-2}, 10^{-4}, 10^{-6}$ (see legend in panel (b)
for associating different dot styles with particular $\tilde i$). 
\label{fig:hat_e}}
\end{figure}

The orbits used in the Monte-Carlo evaluation of integrals have been 
drawn from the distribution of initial orbital parameters appropriate 
for each particular scattering coefficient. When computing 
$\langle\tilde {\bf e}\cdot\Delta\tilde {\bf e}\rangle_{\omega,\tau}$, 
$\langle(\Delta\tilde {\bf e})^2\rangle_{\omega,\tau}$, etc. we
select a set of values of $\tilde e$, $\tilde i$, and $\tilde h$,  
and draw $\tau$ and $\omega$ randomly from a uniform distribution 
between $0$ and $2\pi$. When computing $\hat H_{1,2}$, $\hat K_{1,2}$
we also draw $\tilde h$ from a uniform distribution between $-L_h$
and $L_h$, while keeping $\tilde e$, $\tilde i$ fixed. Finally,
to compute $H_{1,2}$, $K_{1,2}$ we draw $\tilde e_x$, $\tilde e_y$, 
$\tilde i_x$, $\tilde i_y$ randomly 
from a Gaussian distribution (\ref{eq:Gauss}) with given dispersions
$\tilde\sigma_e$ and $\tilde\sigma_i$, while $\tilde h$ is drawn from
a uniform distribution  between $-L_h$ and $L_h$. 

In all our of calculations of $\hat H_{1,2}$, $\hat K_{1,2}$ we use
\ba
L_h=10+4\tilde e,
\label{eq:L_h}
\ea
to ensure that even orbits with $\tilde h>\tilde e$ are properly accounted 
for. When computing $H_{1,2}$, $K_{1,2}$ we use the same prescription
for $L_h$ but with $\tilde \sigma_e$ replacing $\tilde e$. We adopted the 
following prescription for the number of orbits used for evaluating 
scattering coefficients: 
\ba
N=5\times 10^4\left[10+(1+\tilde i)(1+\tilde e)^2\right].
\label{eq:N}
\ea
Thus, to compute scattering coefficients for 
$\tilde e=15$ we run around 13 million scattering calculations.
For $H_{1,2}$, $K_{1,2}$ we used the same prescription 
with $\tilde \sigma_e$, $\tilde \sigma_i$ replacing  $\tilde e$, 
$\tilde i$. 

\begin{figure}
\plotone{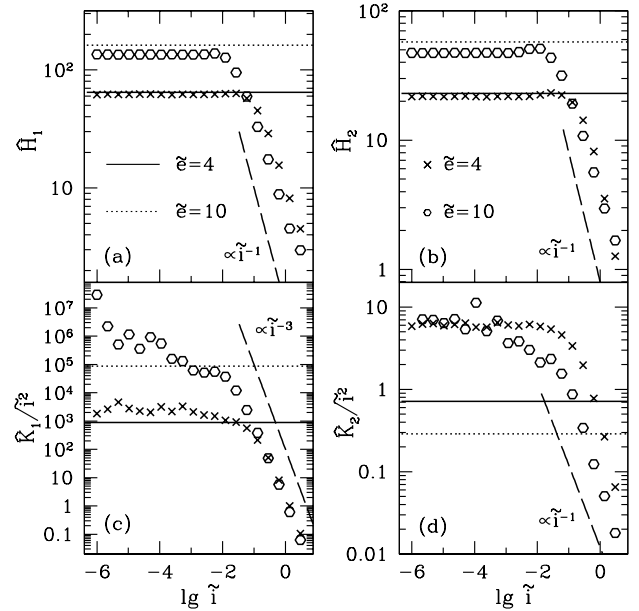}
\caption{
Plots of the same scattering coefficients as in Figure \ref{fig:hat_e} 
but now as a function of $\tilde i$ for $\tilde e=4$ and $\tilde e=10$ 
(see legend in panel (b)). Analytical results for the case of a 
thin disk are shown as solid lines for $\tilde e=4$ and as dotted lines 
for $\tilde e=10$. Long-dashed lines show analytical scaling of 
scattering coefficients in 3D regime, when $\tilde i\gtrsim\tilde e^{-2}$. 
\label{fig:hat_i}}
\end{figure}

In Figure \ref{fig:hat_e} we present the results of calculation 
of $\hat H_{1,2}$, $\hat K_{1,2}/\tilde i^2$ as a function of $\tilde e$,
for several values of $\tilde i=10^{-1}, 10^{-2}, 
10^{-4}, 10^{-6}$, together with our analytical predictions 
(\ref{eq:hatH_1})-(\ref{eq:hatK_2}) shown by solid lines. 
Values of $\hat K_{1,2}$ are scaled by $\tilde i^2$ to
simplify comparison of curves corresponding to different $\tilde i$.

Figure \ref{fig:hat_e}a demonstrates rather good agreement between 
analytical and numerical results for the gravitational stirring 
coefficient $\hat H_1$ almost everywhere in the considered 
range. Agreement at very small values of $\tilde e$ is likely a
coincidence since at $\tilde e\approx 1$ a transition to a shear-dominated 
scattering should occur which invalidates our assumption of 
$\tilde e\gg 1$. At higher values of $\tilde e$ curves corresponding 
to different $\tilde i$ generally line up with the analytical results quite 
well except for the noticeable deviation of $\tilde i=0.1$ curve from 
analytical result (\ref{eq:hatH_1}) which starts around 
$\tilde e\approx 3$ and becomes stronger as $\tilde e$ grows. This deviation
is expected since our analytical results are 
valid only for $\tilde i$ satisfying the constraint (\ref{eq:condit}).
For $\tilde i=0.1$ this means that agreement with the analytical result is 
expected only for $\tilde e\lesssim \tilde i^{-1/2}
\approx 3$, in good correspondence with Figure \ref{fig:hat_e}. 
This point is reinforced by observing the
deviation of $\tilde i=10^{-2}$ results from the analytical curve that 
starts at $\tilde e\approx 10\approx (10^{-2})^{1/2}$, i.e. again 
agrees with the constraint 
(\ref{eq:condit}). Points for $\tilde i=10^{-4}$ and $10^{-6}$ fall
on top of each other in the whole range of calculation as they should.
They lie somewhat below the analytical prediction for large values of 
$\tilde e$, which we do not have a good explanation for. 

The same applies very well to the results for the gravitational friction 
coefficient $\hat H_2$ shown
in Figure \ref{fig:hat_e}b. The only slight difference is that the
influence of the shear-dominated regime is more pronounced for this
scattering coefficient, as $\hat H_2$ settles onto the analytical result
(\ref{eq:hatH_2}) only at $\tilde e\approx 2.5$. Thus, the scattering 
coefficients based on eccentricity changes agree with theory quite 
well within the range of applicability of the analytical results.

\begin{figure}
\plotone{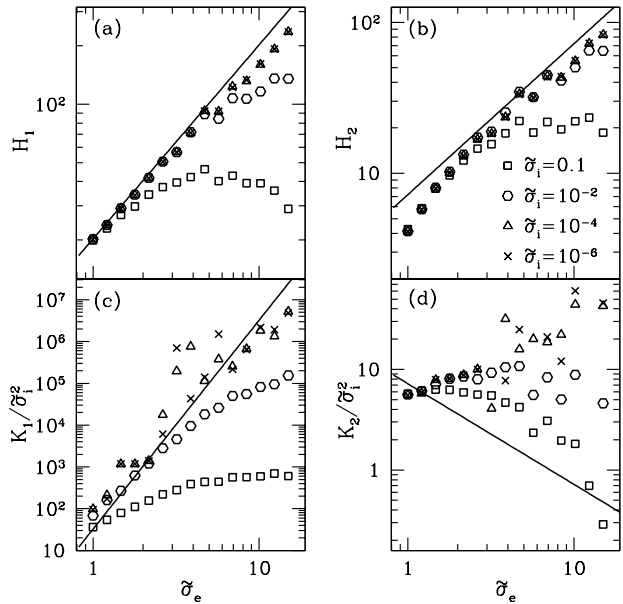}
\caption{
Same as Figure \ref{fig:hat_e} but for scattering coefficients (a) $H_1$,
(b) $H_2$, (c) $K_1/\tilde \sigma_i^2$, (d) $K_2/\tilde \sigma_i^2$, 
corresponding to the Gaussian distribution of $\tilde e$ and $\tilde i$.
Coefficients are shown as a function of $\tilde \sigma_e$ for several
values of $\tilde \sigma_i=10^{-1}, 10^{-2}, 10^{-4}, 10^{-6}$ 
(see legend in panel (b)
for associating different dot styles with different $\tilde \sigma_i$)
\label{fig:gauss_se}}
\end{figure}

We now turn to coefficients $\hat K_{1,2}$ which are based on changes in 
inclination. As one can see from Figure \ref{fig:hat_e}c, the shear-dominated 
regime affects stirring coefficient $\hat K_1$ for $\tilde e\lesssim 2.5$. 
As expected from our previous 
discussion, $\hat K_1$ strongly deviates from the analytical
prediction starting at around $\tilde e\approx 3$ for $\tilde i=0.1$ and 
at around $\tilde e\approx 10$ for $\tilde i=10^{-2}$. However, the results
for $\tilde i=10^{-4}$ and $\tilde i=10^{-6}$  generally do not fall on 
top of each other as one would expect given that the quadratic scaling of 
$\hat K_{1,2}$ with $\tilde i$ has been removed in Figures 
\ref{fig:hat_e}c,d. Moreover, the values of $\hat K_1$ clearly deviate quite
strongly from the analytical prediction (\ref{eq:hatK_1}), sometimes by 
three orders of magnitude, without any recognizable regular pattern.

The numerical results for the gravitational friction coefficient $K_2$
as compared with theory are even more surprising, as Figure 
\ref{fig:hat_e}d demonstrates. Here, for
small values of $\tilde e\lesssim 5$, $\hat K_2$ systematically 
{\it increases} with $\tilde e$, while analytical result (\ref{eq:hatK_2}) 
predicts that $\hat K_2$ should be a {\it decreasing} function of 
$\tilde e$. At larger $\tilde e$ numerically computed values of 
$\hat K_2$ exhibit significant scatter in a chaotic fashion. To be 
fair, one should note that some of the theoretical expectations 
are confirmed by numerics  even for 
$\hat K_2$: a curve for $\tilde i=0.1$ again diverges from  
other curves corresponding to smaller values of $\tilde i$ at 
$\tilde e\approx 3$.

\begin{figure}
\plotone{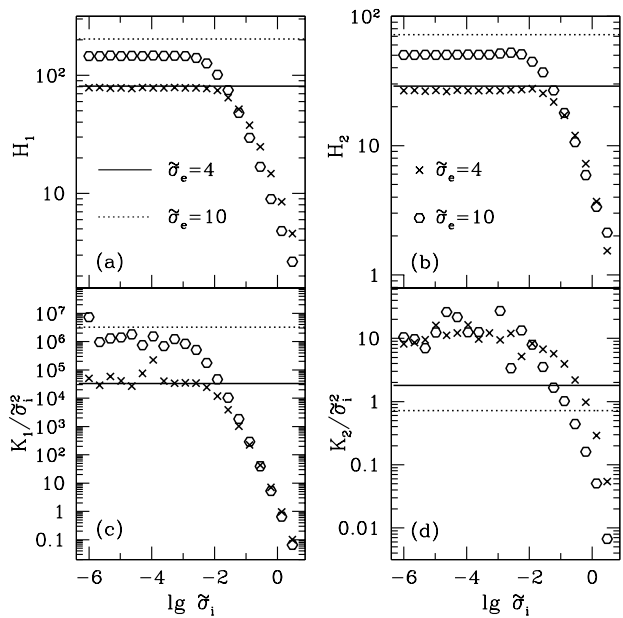}
\caption{
Plots of the same scattering coefficients as in Figure \ref{fig:gauss_se} 
as a function of $\tilde \sigma_i$ for $\tilde \sigma_e=4$ and 
$\tilde \sigma_e=10$ 
(see legend in panel (b)). Analytical results for the case of
thin disk are shown as solid lines for $\tilde \sigma_e=4$ and as dotted lines 
for $\tilde \sigma_e=10$.  
\label{fig:gauss_si}}
\end{figure}

In Figure \ref{fig:hat_i} we look at the behavior of scattering 
coefficients as functions of $\tilde i$ for a fixed value of 
$\tilde e$. The main goal of these plots is to illustrate the transition 
between the thin and thick disk regimes of planetesimal scattering
occuring at $i_{crit}\approx \tilde e^{-2}$. The rather good accuracy 
of our analytical results can be clearly seen in the behavior of 
$\hat H_{1,2}$ and even $\hat K_1$ (the situation is less clear in the case 
of $\hat K_2$): the behavior of the scattering coefficients 
changes dramatically 
at $\tilde i\approx 10^{-2}$ for $\tilde e=10$ and at 
$\tilde i\approx 0.05$ for $\tilde e=4$. The long-dashed lines in 
Figure \ref{fig:hat_i} illustrate the scaling of the scattering coefficients
with $\tilde i$ in the thick-disk regime (Stewart \& Ida 2000), and 
show good agreement with our numerical results when the condition 
(\ref{eq:condit}) is violated. It is clear from Figures 
\ref{fig:hat_i}a,b that our thin-disk theory describes the behavior of
eccentricity-based coefficients  
$\hat H_{1,2}$ quite accurately even for $\tilde e=4$, which is not very 
far from the shear-dominated regime. 

However, from  
Figures \ref{fig:hat_i}c,d one sees once again that the inclination-based 
coefficients $\hat K_{1,2}$ deviate from analytical predictions. Already for 
$\tilde e=4$ 
coefficient $\hat K_1$ exhibits stochastic variations as a function of 
$\tilde i$ by a factor of order unity. At $\tilde e=10$ these variations
become quite dramatic and exhibit an increasing trend with decreasing 
$\tilde i$. This is rather surprising since one expects analytical theory
to work better for very small values of $\tilde i$, when the condition 
(\ref{eq:condit}) is satisfied by a large margin. This clearly indicates 
that the theory is missing some important ingredient, a conclusion which is
additionally reinforced by Figure \ref{fig:hat_i}d demonstrating rather
poor agreement between numerical and analytical values of $\hat K_2$. 

In Figures \ref{fig:gauss_se}, \ref{fig:gauss_si} we show the behavior
of scattering coefficients $H_{1,2}$, $K_{1,2}$ averaged over the Gaussian 
distribution of $\tilde e$ and $\tilde i$. As expected, all the major 
features of $\hat H_{1,2}$, $\hat K_{1,2}$ discussed above are preserved
in these plots, although the overall agreement with theory is additionally
spoiled by the fact that numerically computed $H_{1,2}$, $K_{1,2}$ 
represent a Gaussian convolution of $\hat H_{1,2}$, $\hat K_{1,2}$ over an 
{\it extended} range in $\tilde e$ and $\tilde i$, and not everywhere 
inside this range are the basic assumptions (e.g. dispersion-dominated 
scattering) of our analytical theory 
fulfilled. In particular, numerical coefficients are affected to some extent
by shear-dominated scattering events, not accounted for 
in our theory. Also, at high $\tilde \sigma_i\sim 0.1-10^{-2}$ a 
significant fraction of numerically integrated scattering events 
had values of $\tilde i\sim \tilde e$ corresponding to thick
disk scattering, for which the behavior of coefficients is different
from our theory; see Figure \ref{fig:hat_i}. 

Based on the results presented in Figures \ref{fig:hat_e}-\ref{fig:gauss_si}
we conclude that analytical theory explains quite well the behavior of
scattering coefficients based on changes of $\tilde e$, while it provides 
a rather poor fit to the numerically determined behavior of the 
inclination-based scattering coefficients. There may be several reasons 
for this discrepancy, some of which are listed below.

\begin{enumerate}

\item It may be that the discrepancy arises when we integrate the 
phase-averaged coefficients $\langle\tilde {\bf e}\cdot\Delta\tilde 
{\bf e}\rangle_{\omega,\tau}$, 
$\langle(\Delta\tilde {\bf e})^2\rangle_{\omega,\tau}$, etc. over 
$\tilde h$ to obtain $\hat H_{1,2}$, etc.; see definitions 
(\ref{eq:stirring_coeffs1}) and (\ref{eq:stirring_coeffs2}). In particular,
encounters with $\tilde h>\tilde e$ neglected in our analytical work may 
provide an important contribution to the numerically computed rates.

\item The two-body approximation used in our analytical calculations 
does not work well.

\item Our assumption of a single close scattering per approaching orbit
may be faulty, as the scattered planetesimals may have orbital parameters 
allowing them to experience additional close approaches with the scatterer. 

\item Our theory assumes that the changes in planetesimal orbital
elements occur only during the close approach, when the planetesimal 
separation is $\lesssim R_H$, while in reality it may be that the more
distant interactions between planetesimals at separations $\gtrsim R_H$ 
also play an important role.

\end{enumerate}

We devote the rest of this section to exploring these possibilities.

\begin{figure}
\plotone{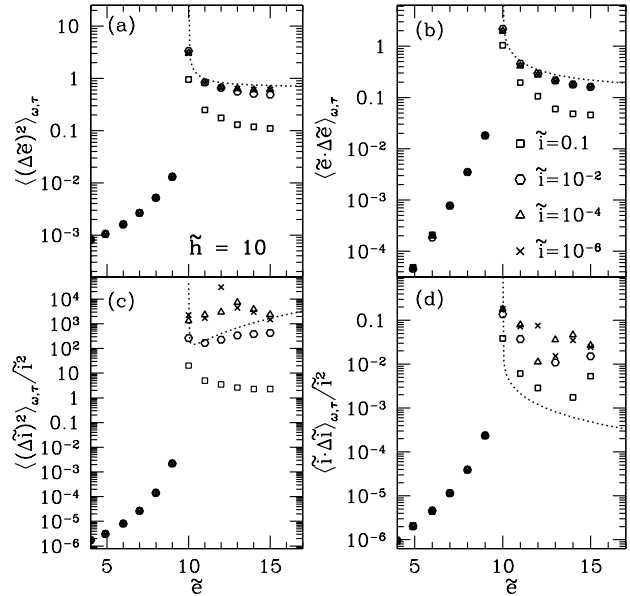}
\caption{
Plots of phase-averaged scattering coefficients (a) 
$\langle(\Delta\tilde {\bf e})^2\rangle_{\omega,\tau}$,
(b) $\langle\tilde {\bf e}\cdot\Delta\tilde 
{\bf e}\rangle_{\omega,\tau}$, (c) 
$\langle(\Delta\tilde {\bf i})^2\rangle_{\omega,\tau}$, (d) 
$\langle\tilde {\bf i}\cdot\Delta\tilde {\bf i}\rangle_{\omega,\tau}$,
as functions of $\tilde e$ for a fixed value of $\tilde h=10$ and 
several values of $\tilde i=0.1,10^{-2},10^{-4},10^{-6}$; see legend 
in panel (b). Note the rapid decay of scattering coefficients
for $\tilde e<\tilde h$. Dotted lines show analytical predictions
for $\tilde e>\tilde h$.
\label{fig:e_var}}
\end{figure}

\subsection{Integration over $\tilde h$.}
\label{ssect:non_close}

\begin{figure}
\plotone{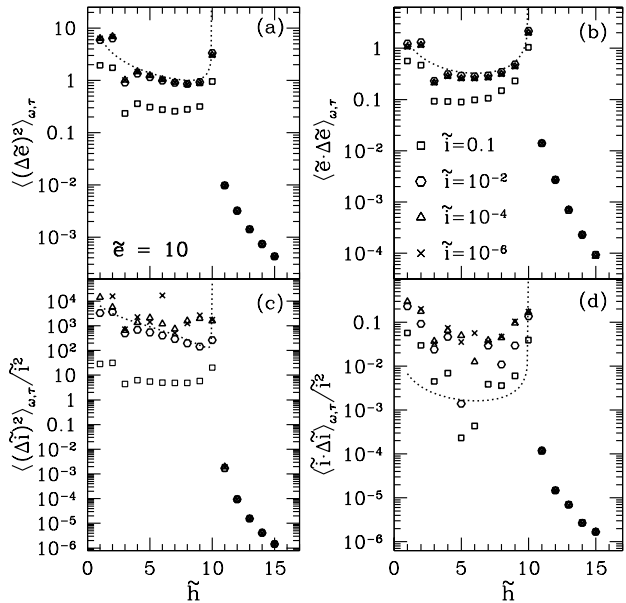}
\caption{
Same as Figure \ref{fig:h_var} but with phase averaged coefficients 
plotted as functions of $\tilde h$ for $\tilde e=10$.
\label{fig:h_var}}
\end{figure}

To figure out whether the aforementioned discrepancy between the analytical 
and numerical inclination-based scattering coefficients can be caused by
the integration of the phase-averaged coefficients over $\tilde h$ we 
look at the behavior of the phase-averaged coefficients. In Figure
\ref{fig:e_var} we present their scaling with $\tilde e$ for a 
fixed value of $\tilde h=10$ and several values of $\tilde i$. Similarly,
in Figure \ref{fig:h_var} these coefficients are shown as functions of 
$\tilde h$ for fixed $\tilde e$ and the same values of $\tilde i$. 
Based on these plots we can make several conclusions.

First, when $\tilde e>\tilde h$ analytical predictions for $\langle\tilde 
{\bf e}\cdot\Delta\tilde {\bf e}\rangle_{\omega,\tau}$ and 
$\langle(\Delta\tilde {\bf e})^2\rangle_{\omega,\tau}$ fit 
our numerical results quite well. Coefficients computed for 
$\tilde i=0.1$ deviate from theory because, as previously described, 
they do not correspond to the thin disk scattering regime.
At the same time $\langle\tilde {\bf i}\cdot\Delta\tilde 
{\bf i}\rangle_{\omega,\tau}$ and 
$\langle(\Delta\tilde {\bf i})^2\rangle_{\omega,\tau}$ are still 
significantly different from theory and exhibit rather erratic 
behavior. 

Second, the values of all scattering coefficients corresponding to 
$\tilde e<\tilde h$ are much smaller than their values for 
$\tilde e>\tilde h$. In the latter case planetesimals can 
experience a close approach, while in the former this is not possible, and 
changes of orbital elements are much weaker than in the latter case.
As a result, contribution of orbits with $\tilde e<\tilde h$ to
scattering coefficients is very small. In fact, one
can see from Figure \ref{fig:h_var} that for $\tilde e=10$,
$\langle(\Delta\tilde {\bf e})^2\rangle_{\omega,\tau}$ 
computed at $\tilde h=10$ (close encounters 
possible) and  $\tilde h=11$ (close encounters not possible)
differ by more than 2 orders of magnitude. The same is true for 
$\langle\tilde {\bf e}\cdot\Delta\tilde {\bf e}
\rangle_{\omega,\tau}$, while for 
$\langle(\Delta\tilde {\bf i})^2\rangle_{\omega,\tau}$ and 
$\langle\tilde {\bf i}\cdot\Delta\tilde {\bf i}
\rangle_{\omega,\tau}$ this difference is 6 and 3 orders of 
magnitude respectively. This very well illustrates our point 
made in \S \ref{sect:scat_coef} that trajectories experiencing 
large-angle scattering (possible only for $\tilde e>\tilde h$)
strongly dominate scattering coefficients in the case 
of a thin planetesimal disk.

To summarize, the results displayed in Figures 
\ref{fig:e_var}-\ref{fig:h_var} make it clear that the well-separated
orbits with $\tilde e<\tilde h$, for which strong scattering is
impossible, do not contribute much to the scattering coefficients.
All the discrepancy between the theoretical and numerical values of 
$\hat K_{1,2}$ and $K_{1,2}$ is already present in the corresponding 
phase-averaged coefficients, and is not introduced by the integration
of the phase-averaged coefficients over only a finite range of 
$\tilde h$, $|\tilde h|<\tilde e$.

\subsection{Accuracy of the two-body approximation.}
\label{ssect:2body}

Next we consider whether the two two-body approximation adopted in our 
analytical calculations is valid for the case of thin disk scattering.

Previously, Tanaka \& Ida (1996) have compared numerically computed 
changes of orbital parameters resulting from gravitational scattering 
with analytical predictions derived in the two-body approximation. 
They found good agreement between the two, except for the narrow regions of 
the initial epicyclic phases in which orbital parameters evolved in
a chaotic manner, if (a) $\tilde h\gtrsim 2$,
(b) the encounter velocity $\tilde v_0\gtrsim 4$ and (c) a small shift 
in the initial 
epicyclic phases $\tau$ and $\omega$ is introduced to match analytical
predictions. Tanaka \& Ida (1996) have compared only 
$\Delta h$ calculated by both methods for $\tilde i=0$, and also the changes 
of other orbital elements for $\tilde i\sim\tilde e\gtrsim 1$. None
of these cases corresponds to the regime of thin disk scattering 
considered in this work although the former does describe quite well
the variation of the eccentricity-based scattering coefficients. For
that reason we ran our own calculations with initial orbital parameters
selected to correspond to the thin disk case.

\begin{figure}
\plotone{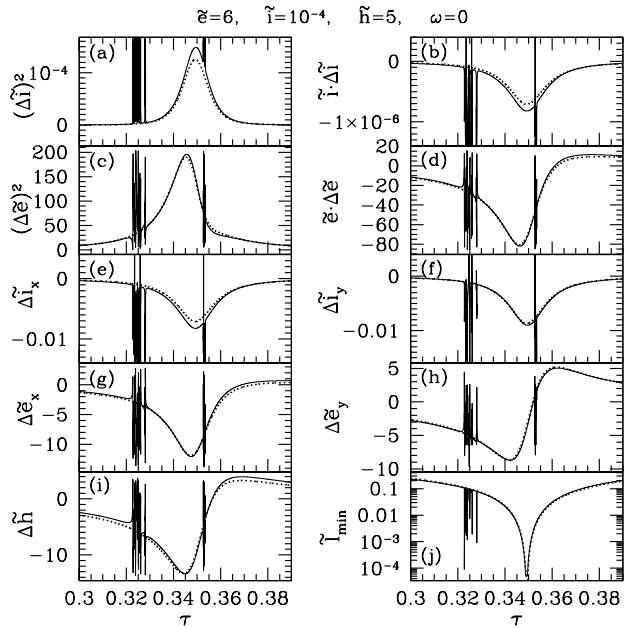}
\caption{
Changes of relative orbital parameters  (a) $(\Delta \tilde i)^2$, 
(b) $\tilde i\cdot \Delta \tilde i$, (c) $(\Delta \tilde e)^2$, (d)
$\tilde e\cdot \Delta \tilde e$, (e) $\Delta \tilde i_x$, 
(f) $\Delta \tilde i_y$, (g) $\Delta \tilde e_x$, (h) $\Delta \tilde e_y$,  
(i) $\Delta \tilde h$, and (j) the 
minimum separation $\tilde l_{min}$ of planetesimals plotted as 
a function of their relative horizontal epicyclic phase $\tau$ (at large
separation prior to scattering). Initial orbital parameters corresponding
to this calculation are shown at the top of the plot. 
Only an interval $0.3<\tau<0.39$ in which
strong scattering takes place is displayed. Solid curves show the numerical
results while the dotted lines are the analytical predictions. Note
the chaotic variation of orbital parameters for  
$0.32<\tau<0.33$ and $0.35<\tau<0.355$.
\label{fig:pars_of_tau}}
\end{figure}

In general we find good agreement with the conclusions of Tanaka 
\& Ida (1996), as shown in particular in Figure \ref{fig:pars_of_tau}
where we display the changes of various orbital elements resulting from 
gravitational scattering as well as the minimum approach distance between 
the scattering bodies $\tilde l_{min}$. In making this Figure we have 
slightly shifted analytical curves (shown as dotted lines) in 
$\tau$ by $\Delta \tau=-0.05$ to
make them better match numerical results (shown as solid curves). 
In practice such a shift
of the orbital phase arises due to the distant interaction between 
the planetesimals as they approach each other (Tanaka \& Ida 1996).
Only one interval of $\tau$ in which strong scattering is possible
is shown, $0.3<\tau<0.39$; another one exists at $5.97<\tau<6.05$, 
in accordance with the discussion in Appendix, where the existence 
of two values of $\tau$ for which close encounters are possible
for $\tilde e>\tilde h$ is stated.

One can deduce from Figure \ref{fig:pars_of_tau} that analytical 
curves follow the numerical results quite well for the majority of 
values of $\tau$ except for the two narrow ranges of $\tau$, namely 
$0.32<\tau<0.33$ and $0.35<\tau<0.355$. Inside these intervals orbital
elements experience strong chaotic variations as $\tau$ changes,
with $\Delta \tilde h$, $\Delta \tilde e_x$, $\Delta \tilde e_y$
deviating from analytical prediction by a factor of order unity, while 
$\Delta \tilde i_x$, $\Delta \tilde i_y$ differ from theory by 
several orders of magnitude (off scale on these plots)! Although 
these deviations are very significant we will show next that they
are not caused by the failure of the two-body approximation. 
Thus, use of the two-body approximation cannot explain the discrepancy
between the numerical and analytical inclination-based scattering 
coefficients.

\subsection{Single-scattering approximation.}
\label{ssect:single_scat}

Our calculation has always assumed that changes of orbital elements 
arising during a  scattering event are
final. In reality there may be a situation when a post-scattering 
orbital elements are such that they cause another close approach between 
planetesimals, leading to additional variation of orbital elements. And
this may happen not just once for a given incoming orbit. Such multiple 
scattering events are very typical for planetesimals scattering in the 
shear-dominated regime but their importance in the high-velocity case
is not very obvious.

To see that multiple scattering is indeed possible even in the 
dispersion-dominated regime we take a closer look at Figure 
\ref{fig:pars_of_tau}i where we plot $\Delta \tilde h$ as a function 
of $\tau$. One can see that chaotic orbits strongly deviating from 
analytical prediction (shown as a dotted line) exist almost solely in 
those regions where $\Delta h$ predicted by theory happens to be 
$\approx -\tilde h$ ($\Delta h\approx -5\approx -\tilde h$ in the case 
displayed in this Figure). This is not a coincidence, and what really 
happens is the following. First, planetesimals scatter and their 
orbital elements change in full agreement with analytical theory.
This means, however, that post-scattering $\tilde h\sim 1$ and the
guiding center of the planetesimal orbit is now moving very slowly 
with respect to the scatterer, while eccentricity is still quite high.
Right after scattering planetesimals are very close to each other,
and Keplerian shear does not allow their guiding centers to recede 
very far  because $\tilde h$ is small, so that after one orbital 
period planetesimals may closely approach each other again and 
experience another scattering. This second scattering may or may not 
dislodge them from close proximity of each other but it will certainly 
affect their final orbital elements, explaining the deviation of
$\Delta \tilde h$ and other orbital elements from theoretical 
predictions when $\Delta h\approx -\tilde h$. Thus, we conclude that 

\begin{itemize}

\item Chaotic variations of 
orbital elements result from multiple scatterings of planetesimals 
in the course of close encounter rather than from the failure of 
the two-body approximation.

\item Phase intervals where theoretical 
$\Delta \tilde h\approx -\tilde h$ are naturally occupied by 
chaotic orbits (there are also other possibilities for producing 
multiple scattering orbits, see below).

\end{itemize}

These observations greatly help in explaining the puzzling results 
for the inclination-based scattering coefficients. From pure geometry
it is clear that the highest inclination $\tilde i_1$ which a 
high-velocity particle with initial inclination $\tilde i_0$ can attain 
after a single scattering event is\footnote{Highest inclination 
results from scattering by $\approx \pi/2$, which requires impact 
parameter of incoming trajectory to be $\tilde l\sim \tilde v_0^{-2}$,
at initial vertical separation of order $\tilde i_0 R_H$. 
The final velocity of the receding planetesimal is $\tilde v_0$ and 
from simple geometry its vertical component (which is equivalent to 
inclination in Hill units) is $\tilde v_0\times (\tilde i_0/\tilde l)
\sim \tilde i_0\tilde v_0^3$.} 
\ba
\tilde i_1\sim \tilde i_0\tilde v_0^3. 
\label{eq:i1}
\ea
This is easily seen in 
Figures \ref{fig:pars_of_tau}e,f which show that maximum 
$\tilde i_1\sim 10^{-2}$ for $\tilde i_0\sim 10^{-4}$: since 
$\tilde v_0=(\tilde e^2-(3/4)\tilde h^2)^{1/2}
\approx 4.2$ one should expect maximum $\tilde i_1\sim 10^{-4}\times
4.2^3\approx 0.007$, very close to what we find in reality
outside of the region of chaotic orbits. 

At the same time Figures \ref{fig:pars_of_tau}e,f show that chaotic 
orbits often exhibit final $\tilde i$ much larger than predicted by 
equation (\ref{eq:i1}). This, of course, is naturally explained by 
the fact that chaotic orbits result from {\it multiple} scattering. 
Every scattering of a high-velocity orbit can potentially increase
inclination by a factor $\tilde v_0^3\gg 1$ (the approach velocity 
of planetesimals does not change very strongly after multiple 
scatterings and is still $\sim \tilde v_0$; this can be understood 
from the conservation of Jacobi constant). Thus, after $n$ scatterings
maximum possible inclination would have been $\tilde i_n\sim \tilde i_0
\tilde v_0^{3n}$, except that in practice $\tilde i$ cannot exceed 
$\tilde v_0$. The highest final $\tilde i$ that we could 
find for the parameters of Figure \ref{fig:pars_of_tau} is $\approx 0.7$ but 
one has to keep in mind that orbits in the chaotic region exhibit 
quasi-fractal behavior in that the denser is the grid in $\tau$ used for 
computing $\Delta \tilde i$ the richer the behavior found. Thus, 
we could have easily missed orbits with even higher final $\tilde i$. On
theoretical grounds we expect the maximum $\tilde i$ in this Figure at the 
level of $10^{-4}\times 4.2^{3\times 2}\approx 0.5$ if only $n=2$ 
close scatterings have taken place.

\begin{figure}
\plotone{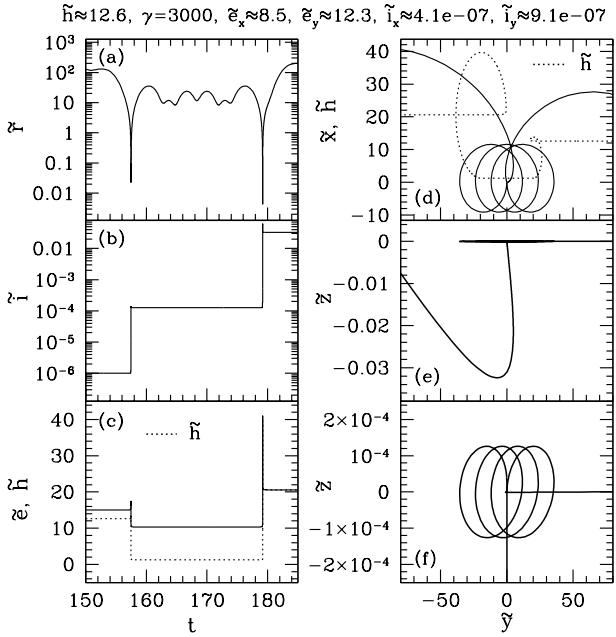}
\caption{
(a-c) Variation of the relative orbital elements of two planetesimals in 
the course of a multiple scattering event, for initial orbital elements 
indicated at the top of the plot. Evolution of (a) the relative distance 
between the bodies $\tilde r$, (b) their relative inclination $\tilde i$,
and (c) their $\tilde e$ (solid line) and $\tilde h$ (dotted line) are 
shown. This trajectory exhibits two strong scattering events in each of 
which $\tilde i$ gets boosted up by two orders of magnitude. 
(d-f) Trajectory of relative motion in the course of scattering
shown (d) in the $\tilde y-\tilde x$ coordinates and (e) in 
$\tilde y-\tilde z$ coordinates. A zoomed in version of panel (e) is
shown in panel (f) to better illustrate the complexity of vertical motion 
before the final scattering causes planetesimals to recede from each other.
In panel (d) we show both the instantaneous position of planetesimal that
is being scattered (solid line) and the trajectory of its guiding
center (dotted line). See text for more details.
\label{fig:orb4}}
\end{figure}

In Figure \ref{fig:orb4} we illustrate a multiple scattering event for
an orbit with initial parameters $\tilde e=15, \tilde i=10^{-6}$ and 
$\tilde h\approx 12.6$ (in this case $\tilde v_0\approx 10.2$). From Figure 
\ref{fig:orb4}b,c one can see that as a result of the first scattering 
$\tilde h$ becomes very small while $\tilde i$ jumps up by $\sim 10^2$.
After that the planetesimal loops around its scatterer for several orbital 
periods, as illustrated in Figures \ref{fig:orb4}d-f, until the second 
strong scattering takes place, resulting in final $\tilde h\approx 20$. 
This allows the planetesimal to leave the vicinity of its scatterer.
During the second scattering $\tilde i$ is again boosted up by more than 
two orders of magnitude, resulting in final $\tilde i\approx 0.03$. This 
is much larger than $10^{-6}\times 10.2^3\approx 10^{-3}$ --- the maximum 
$\tilde i_1$ one would expect from single scattering.

It now becomes much easier to understand the erratic behavior of 
scattering coefficients $\hat K_{1,2}$ in Figure \ref{fig:hat_e}. 
In particular, multiple scattering orbits very strongly affect 
$\hat K_1$ since $\Delta \tilde i$ enters the 
calculation of this stirring coefficient in a second power. Because of 
that, even though chaotic 
orbits arise for only a small subset of horizontal epicyclic phases they
affect the value of numerically determined $\hat K_1$ very strongly. 
According to equation (\ref{eq:di2_omtau}) 
$\langle (\Delta \tilde i)^2\rangle_{\omega,\tau}\approx 6.5\times 
10^{-10}$ for the values of $\tilde e, \tilde h$
and $\tilde i$ used in making Figure \ref{fig:orb4}. At the same time
a single orbit as displayed in Figure \ref{fig:orb4} has 
$(\Delta \tilde i)^2\sim 10^{-4}$, more than $10^6$ times higher than 
the theoretical phase average of this quantity. It is thus not 
surprising that even though we have used a very large number of orbits
in computing scattering coefficients (according to the prescription 
(\ref{eq:N}) our calculation of $\hat K_1$ for $\tilde e=15$ used 
13.3 million orbits) chaotic orbits still affect them quite 
significantly. 

We can now explain why the scatter in numerical values of $\hat K_1$ 
and the deviation from analytical prediction both become stronger 
as $\tilde i$ decreases: the maximum possible value of $\tilde i$ 
resulting from scattering
is always limited from above by $\tilde i\sim \tilde v_0$, so that the 
maximum stochastic $(\Delta\tilde i)^2\sim v_0^2$, independent 
of initial $\tilde i$. However, the analytical 
value of $\hat K_1\propto\tilde i^2$, so that the ratio of analytical
$\hat K_1$ to the numerical one  increases
as $\tilde i$ decreases. 

It is not even clear that our calculation of $\hat K_1$ in Figure 
\ref{fig:hat_e} has
converged --- one cannot guarantee that increasing the number of orbits 
would not increase even more the number of extremely chaotic orbits with
very large $\Delta \tilde i$, which would then dominate the calculation. 
The only thing that argues against this scenario is the saturation of
final $\tilde i$ at the level of $\tilde v_0$ even for very large number of
repeated scatterings. Nevertheless, until we understand how much of the 
phase space volume corresponds to chaotic orbits with very large 
$\Delta \tilde i$
we cannot draw a final conclusion about the convergence of $\hat K_1$ and 
we leave this subject for future investigation. Paradoxically, the 
agreement between the numerical and analytical results may be better 
if one uses smaller number of orbits in numerical calculation of 
$\hat K_{1,2}$ since then the chance of randomly picking one of the
high-$\Delta \tilde i$, multiple scattering orbits is also smaller.

\begin{figure}
\plotone{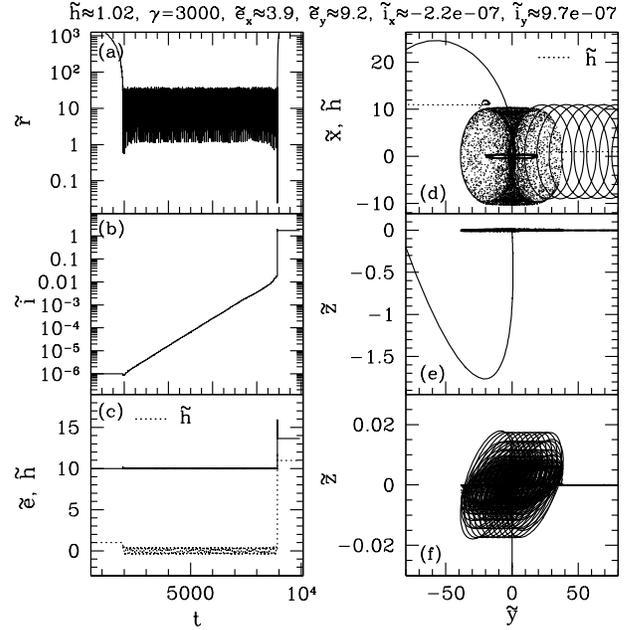}
\caption{
Same as Figure \ref{fig:orb4} but for a different choice of orbital 
elements indicated at the top of the plot characterized by small 
initial $\tilde h$. Note an exponential growth 
of $\tilde i$ by several orders of magnitude in panel (b). In panel 
(d) we plot only a small fraction of data during the scattering event 
as dots to better illustrate the underlying structure. See text 
for more details.
\label{fig:orb1}}
\end{figure}

\begin{figure}[t]
\plotone{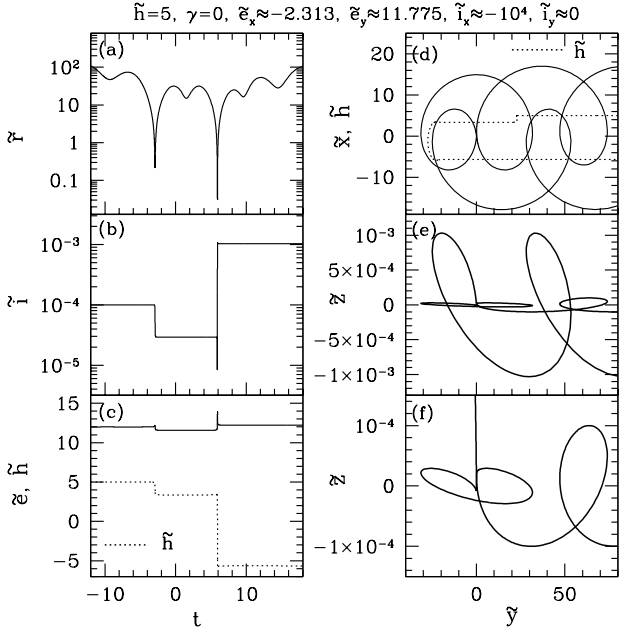}
\caption{
Same as Figure \ref{fig:orb1} but for a different choice of orbital 
elements indicated at the top of the plot. Note that while after the first 
scattering $\tilde h$ is not particularly close to unity, 
multiple scattering still takes place. See text for more details.
\label{fig:orb5}}
\end{figure}

In the course of our investigation we have also found that orbits 
with $\tilde e\gg 1$ and $\tilde h\gg 1$ (like the one shown in 
Figure \ref{fig:orb4}) are not the biggest contributors to chaos
in $\hat K_{1,2}$. It turns out that inclination-based scattering 
coefficients are most strongly affected  by orbits with 
$\tilde e\gg 1$ and $\tilde h\sim 1$, i.e. orbits which are initially 
close to the separatrix between the horseshoe and passing orbits. 
An example of planetesimal scattering corresponding to this case 
is shown in Figure \ref{fig:orb1} for initial $\tilde h\approx 1.02$, 
$\tilde e=10$, $\tilde i=10^{-6}$. This event is characterized by
a very long time interval, more than $10^3$ orbital periods, during 
which planetesimals stay close to each other. They essentially form 
a temporary distant satellite system (note that the distance between
planetesimals is larger than $R_H$), which slowly evolves in time. 
Figure \ref{fig:orb4}b demonstrates that during this temporary capture 
$\tilde h$ oscillates around 
zero not allowing planetesimals ro recede from each other. 
Their relative inclination increases exponentially (with rather long 
time constant) by 4 orders of magnitude in an orderly fashion. Finally a 
strong scattering event occurs, which boosts up $\tilde i$ by $\sim 10^2$ 
and dislodges the planetesimal from its scatterer's vicinity. For this 
event $(\Delta \tilde i)^2\sim 1$, while a single scattering calculation
would predict $\langle (\Delta \tilde i)^2\rangle_{\omega,\tau}
\approx 6.7\times 10^{-9}$ for the values of $\tilde e, \tilde h$
and $\tilde i$ shown on top of Figure \ref{fig:orb5}. As a result,
this single scattering event completely determines the calculation
of $\hat K_1$. 

It is worth pointing out here that multiple scattering in general does not 
require $\Delta \tilde h\approx -\tilde h$ and Figure \ref{fig:orb5}
illustrates this statement. This Figure shows a double scattering event 
for initial $\tilde h=5$, $\tilde e=12$, and $\tilde i=10^{-4}$. As can be 
seen in Figure \ref{fig:orb5}c after the first strong scattering event
$\tilde h\approx 3.5$ and Keplerian shear ensures that the bodies will
not stay close to each other for very long. However, before the planetesimal
leaves its scatterer's vicinity its epicyclic motion brings it back into
their mutual Hill sphere where another scattering event occurs. This type of 
multiple scattering event does not affect coefficients $\hat K_{1,2}$
nearly as much as events with $\Delta \tilde h\approx -\tilde h$.

To summarize, multiple scattering explains the discrepancy between the
analytical and numerical results for $\hat K_1$ and stochastic 
scatter at high $\tilde e$ in values of $\hat K_2$ quite well. However,
this explanation does not work so well for the systematic deviation 
of the numerical $\hat K_2$ from the analytical one clearly seen in Figure 
\ref{fig:hat_e}d for virtually all values of $\tilde e$: even at small
$\tilde e$, when the stochastic scatter is small, $\hat K_2$ 
increases contrary to theory. Note
that while the lower envelope of $\hat K_1$ for a given $\tilde i$ agrees 
quite well with the analytical prediction (and multiple scattering explains 
the remaining stochasticity), this is clearly not true for $\hat K_2$.    

\subsection{Distant interaction.}
\label{ssect:distant_inter}

To understand the systematic deviation of the numerical $\hat K_2$ 
from the analytical prediction (\ref{eq:hatK_2}) we first note that
coefficient $\langle\tilde {\bf i}\cdot\Delta\tilde {\bf i}
\rangle_{\omega,\tau}$ used in calculating of $\hat K_2$  
also systematically differs from the analytical prediction given by
equation (\ref{eq:idi_omtau}); see Figures 
\ref{fig:e_var}d, \ref{fig:h_var}d. This may seem surprising 
since in Figure \ref{fig:pars_of_tau}b the numerically determined 
$\tilde {\bf i}\cdot\Delta\tilde {\bf i}$ follows quite closely 
analytical prediction as a function of horizontal phase $\tau$.
Chaotic variations of $\tilde {\bf i}\cdot\Delta\tilde {\bf i}$ 
due to multiple scattering noticeable in 
this plot within narrow intervals of $\tau$, cannot explain the 
systematic discrepancy for $\hat K_2$ --  they can only be 
responsible for the random scatter in the calculation of $\hat K_2$.
However, Figure \ref{fig:pars_of_tau}b does not show how 
$\tilde {\bf i}\cdot\Delta\tilde {\bf i}$ depends on $\omega$
(this Figure was made for a single value of $\omega$) and this 
dependence turns out to be very important.

\begin{figure}
\plotone{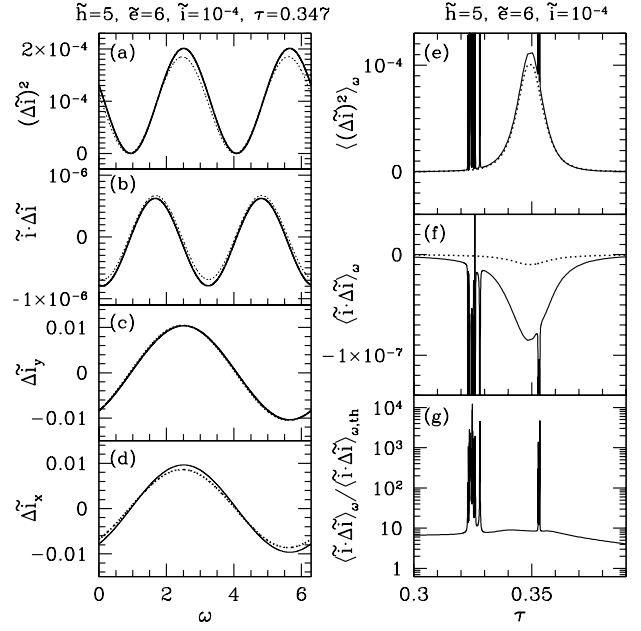}
\caption{
(a-d) Variation of (a) $(\Delta \tilde i)^2$, (b) 
$\tilde i\cdot\Delta \tilde i$, (c) $\Delta \tilde i_y$, and  
(d) $\Delta \tilde i_x$ with vertical epicyclic phase $\omega$
for a particular set of initial orbital parameters (shown on top
of left column). Theoretical predictions are shown with the dotted 
line while numerical results are in solid. (e-g) Plots of 
(e) $(\Delta \tilde i)^2$ and (f) $\tilde i\cdot\Delta \tilde i$
averaged over $\omega$ as functions of $\tau$ for a particular set 
of initial orbital parameters (shown on top of right column).
Solid lines are numerical results, dotted lines are analytical 
predictions. Panel (g) shows the ratio of the numerically computed
average of $\tilde i\cdot\Delta \tilde i$ over $\omega$ to the 
analytical prediction for the same quantity. Panels (f) and (g)
clearly demonstrate that theory underpredicts 
$\langle\tilde i\cdot\Delta \tilde i\rangle_\omega$ by about an 
order of magnitude. 
\label{fig:omega_av}}
\end{figure}

In Figure \ref{fig:omega_av}a-d we display the dependence of
$(\Delta \tilde i)^2$, $\tilde i\cdot\Delta \tilde i$, 
$\Delta \tilde i_y$, and  $\Delta \tilde i_x$ on $\omega$
for a fixed $\tau$. The value of $\tau=0.347$ is chosen in 
order to avoid intervals of chaotic variation of the orbital 
elements\footnote{See Figure \ref{fig:pars_of_tau} which is made
for the same set of initial orbital parameters as Figure 
\ref{fig:omega_av}.} in order to isolate the subsequent analysis 
from the effects of multiple scattering. The general
agreement of the numerical and analytical curves, including those of 
$\tilde i\cdot\Delta \tilde i$, is quite good in this Figure. 
However, to calculate $\langle\tilde {\bf i}\cdot\Delta\tilde 
{\bf i}\rangle_{\omega,\tau}$ we need to {\it integrate}  
$\tilde i\cdot\Delta \tilde i$ over $\omega$ and it is quite
obvious from Figure \ref{fig:omega_av}b that 
$\tilde i\cdot\Delta \tilde i$ is very
close to a pure sinusoid. Its $\omega$-average 
$\langle\tilde {\bf i}\cdot\Delta\tilde 
{\bf i}\rangle_{\omega}$ should then be very small 
and strongly dependent on the {\it deviations} of 
$\tilde i\cdot\Delta \tilde i$ from a pure sinusoid. As a result,
if the numerical and analytical values of 
$\tilde i\cdot\Delta \tilde i$ deviate from the sinusoid {\it differently},
one can get a significant discrepancy between theory and numerical
calculation.

This is exactly what is going on as we demonstrate in Figure 
\ref{fig:omega_av}e,f where we plot $\langle(\Delta \tilde i)^2
\rangle_{\omega}$ and $\langle\tilde {\bf i}\cdot\Delta\tilde 
{\bf i}\rangle_{\omega}$ as functions of $\tau$. One can see that
averaging over $\omega$ does not affect the agreement between 
numerical and analytical $(\Delta \tilde i)^2$ seen in Figure
\ref{fig:omega_av}a because it is an intrinsically positive 
quantity. But $\omega$-averaging of $\tilde i\cdot\Delta \tilde i$
does lead to a dramatic difference between analytical
and numerical $\langle\tilde {\bf i}\cdot\Delta\tilde {\bf i}
\rangle_{\omega}$ in Figure \ref{fig:omega_av}f for the reason 
we just described. In Figure \ref{fig:omega_av}g we show the
ratio of $\langle\tilde {\bf i}\cdot\Delta\tilde {\bf i}
\rangle_{\omega}$ determined by the two methods, 
and one can clearly see that the numerical result significantly
exceeds the analytical one (both in regions of chaotic and orderly 
behavior of orbital parameters), in agreement with the fact that
the numerical $\hat K_2$ is systematically higher than  
the analytical $\hat K_2$, see Figure \ref{fig:hat_e}d.

What may produce such a difference in scaling of analytical and 
numerical $\tilde i\cdot\Delta \tilde i$ with $\omega$? In 
Appendix \ref{app2} we provide a simple calculation of $\langle\tilde 
{\bf i}\cdot\Delta\tilde {\bf i}\rangle_{\omega}$ allowing for
a small but non-zero difference $\delta\omega_{dist}$ of the relative 
vertical epicyclic phase of interacting planetesimals $\omega$ 
at large separation and right before the close encounter. Such a phase
difference in $\omega$ arises because of the distant interaction between
planetesimals prior to their encounter and is analogous to the shift in 
horizontal phase $\tau$ which was invoked in Figure \ref{fig:pars_of_tau} 
to better match analytical and numerical results (see also Tanaka \& 
Ida 1996). We show in Appendix \ref{app2} that although 
$|\delta\omega_{dist}|$ is expected to be small its effect on the 
calculation of $\langle\tilde 
{\bf i}\cdot\Delta\tilde {\bf i}\rangle_{\omega}$ is very important
(and dominates this calculation) as long as 
$\tilde v_0^3|\delta\omega_{dist}|\gtrsim 1$. 

The rather surprising 
result that a small variation of vertical phase $\omega$ can strongly 
affect the calculation of dynamical friction coefficients $\hat K_2$ and
$K_2$ is explained by strong cancellation that takes place when 
one averages $\tilde {\bf i}\cdot\Delta\tilde {\bf i}$ over 
$\omega$: some terms proportional to $\delta\omega_{dist}$ that 
average to zero when $\delta\omega_{dist}\equiv 0$ can become very 
large after averaging when $\delta\omega_{dist}\neq 0$. In graphical 
form the same issue has already been illustrated in Figure
\ref{fig:omega_av}. Also, a careful inspection of our calculation
of all other scattering coefficients including 
$\langle(\Delta \tilde i)^2\rangle_{\omega,\tau}$ shows that
unlike $\langle\tilde 
{\bf i}\cdot\Delta\tilde {\bf i}\rangle_{\omega,\tau}$
they are not affected by non-zero $\delta\omega_{dist}\neq 0$
since they do not suffer from cancellation effects when averaged
over $\omega$. For these coefficients, our calculation 
neglecting the effects of distant interaction presented in 
Appendix \ref{app1} remains valid. 

Another important point to make here is that although there is 
a large difference between the numerical and analytical 
$\hat K_2$ and $K_2$, this discrepancy is not always critical
for the inclination evolution of planetesimals in the regime 
(\ref{eq:condit}). Indeed, as Figures 
\ref{fig:gauss_se}-\ref{fig:gauss_si} show, $K_1/K_2\sim 10^4$ 
for $\tilde \sigma_e\gtrsim 5-10$, which according
to equation (\ref{eq:homog_heating}) implies that the gravitational friction
term becomes important for inclination evolution only if 
$\mu_1\gg\mu_2$, i.e. for the velocity evolution of massive bodies 
driven by their interaction with low-mass planetesimals. 

The issue of distant interaction and its effect on the scattering 
calculation clearly deserves further 
work but we postpone it for a separate investigation. For now it
is enough for us to just state that distant interactions serve 
as a plausible explanation for the deviations between the numerical 
and analytical calculations of 
$\langle\tilde {\bf i}\cdot\Delta\tilde {\bf i}\rangle_{\omega,\tau}$, 
$\hat K_2$ and $K_2$.

%%%%%%%%%%%%%%%%%%%%%%%%%%%%%%%%%%%%%%%%%%%%%%%%%%%%%%%%%%%
\section{Velocity evolution of protoplanetary cores.}
\label{subsect:velev}

Here we use our results as obtained in previous sections to understand 
the velocity evolution of a sparse population of protoplanetary 
cores in the end of oligarchic phase, when their orbits become crossing 
--- a situation described in \S \ref{sect:intro}. 
For simplicity we will assume the masses of all cores $M_c$ to be roughly 
equal, i.e. $\mu_1=\mu_2=\mu=M_c/M_\star$ 
and $\sigma_{\{e,i\}, 1}=\sigma_{\{e,i\}, 2}
=\sigma_{\{e,i\}}$, which allows us to
omit the term proportional to $K_2$ in the equation for inclination 
evolution; see \S \ref{ssect:distant_inter}. Also, we will neglect the 
effects of multiple scattering on the inclination evolution, which 
is justified to some extent by the small number of bodies involved 
(and the small number of their close encounters); see 
\S \ref{ssect:single_scat}. We can then use our analytical 
expressions for the scattering coefficients to study velocity evolution.  

We assume that initially $\tilde \sigma_e=\tilde \sigma_{e 0}\gg 1$ and 
$\tilde \sigma_i=\tilde \sigma_{i 0}$
satisfies constraint (\ref{eq:condit}). Equation (\ref{eq:homog_heating}) 
supplemented with expressions for the scattering coefficients yields the 
following set of velocity evolution equations:
\ba
&& \frac{d\sigma_{e}}{dt}=C_e T_e^{-1}\mu^{1/3},
\label{eq:velev_h}\\
&& \frac{d\sigma_{i}}{dt}=C_i T_e^{-1}\sigma_{i}\tilde \sigma_{e}^5,
\label{eq:velev_v}
\ea
where evolution time $T_e$ is given by
\ba
&& T_e=\frac{1}{\Omega N_p R_H^2}=\frac{\mu^{1/3}M_\star}{\Omega\Sigma_p a^2}
\label{eq:Te}\\
&& \approx 150~\mbox{yr}~
\left(\frac{M_c}{0.01~M_\oplus}\right)^{1/3}
\left(\frac{\rm AU}{a}\right)^{1/2}
\left(\frac{30~ \rm{g ~cm}^{-2}}{\Sigma_p}\right)\nonumber
\ea
for $M_\star=M_\odot$, and 
coefficients $C_e$, $C_i$ are constants of order unity (their
values can be found from equations 
(\ref{eq:H_1})-(\ref{eq:K_1})). Note that these equations apply equally
well both to the dense population of planetesimals with overlapping 
orbits and to the sparse population of cores with crossing orbits. 

One can understand the origin of these equations qualitatively based on the 
fact that scattering coefficients are dominated by the large-angle 
scattering events in the thin-disk case, i.e. those that require impact
parameter $l_{min}\sim R_H/\tilde v_0^2$. Since we consider 
the case $\tilde e\gg 1$ the relative velocity between planetesimals 
is $\sim \tilde v_0\Omega R_H$. Then a mean time between encounters of
a given body with other bodies at an impact parameter $l_{min}$ is
$T_{\pi/2}\sim \tilde v_0/(\Omega N_p R_H^2)$, where $N_p=\Sigma_p/M_c$
is the planetesimal (or core) surface number density. When a large-angle
scattering event occurs $e^2$ changes by $\sim e^2$, while $i^2$
changes by $\sim i^2\tilde v_0^6$, see \S \ref{ssect:single_scat}. Then
$d\sigma_e^2/dt\sim e^2/T_{\pi/2}$, while $d\sigma_i^2/dt\sim i^2
\tilde v_0^6/T_{\pi/2}$, which results in equations 
(\ref{eq:velev_h})-(\ref{eq:velev_v}) if we recall that 
$\tilde v_0\sim \tilde \sigma_e$.

Integrating equations (\ref{eq:velev_h})-(\ref{eq:velev_v}) we find
\ba
&& \tilde \sigma_e=\tilde \sigma_{e 0}+C_e\frac{t}{T_e},
\label{eq:sig_e_sol}\\
&& \tilde \sigma_i=\tilde \sigma_{i 0}\exp\left[\frac{C_i}{6C_e}
\left(\tilde \sigma_e^6-\tilde \sigma_{e 0}^6\right)\right].
\label{eq:sig_i_sol}
\ea
For $t\ll T_e$ one has $\tilde \sigma_e\approx \tilde \sigma_{e 0}$
and 
\ba
\tilde \sigma_i=\tilde \sigma_{i 0}\exp\left[C_i
\tilde \sigma_{e 0}^5\frac{t}{T_e}\right].
\label{eq:sig_i_sol_1}
\ea
One can see from these solutions that growth of $\sigma_i$ has a 
rather explosive character: $\sigma_i$ increases exponentially and 
at the very beginning the inclination growth timescale is 
$\sim \tilde \sigma_{e 0}^{-5} T_e\ll T_e$ since $\sigma_{e 0}\gtrsim 1$.
As a result, while $\sigma_i$ grows by several orders of magnitude,
$\sigma_e$ does not change that much. This means that protoplanetary 
cores should very rapidly transition from the very thin, almost 2D 
configuration to a vertically extended disk in which the condition
(\ref{eq:condit}) is no longer fulfilled. 

In reality inclination will grow not in a continuous fashion as 
described by equations (\ref{eq:sig_i_sol}) and (\ref{eq:sig_i_sol_1})
but more in a step-like way by a factor of $\sim \tilde v_0^3$ 
as large-angle scattering events occur
at time intervals of order $T_{\pi/2}$. Continuous description is going to
be useful only after a large number of large-angle scattering events have
taken place (and during this period $\tilde i$ would have grown a lot). 

Onset of the core orbit crossings not only increases the cores'
inclination but also makes possible collisions between cores leading 
to their growth. While the disk is geometrically thin, the collision 
probability of cores is rather large, and one may wonder whether the cores 
would grow appreciably during the short period of time while the
condition (\ref{eq:condit}) is still fulfilled. What matters for the 
core growth during this period is both the accretion rate (which 
is very high) and the time during which the inclination of the disk is still
in the regime  (\ref{eq:condit}), which is short. Careful analysis shows 
that the relative core mass increase during the ``thin disk'' epoch of 
the core population evolution is very small, much less than unity,
provided that $\tilde v_0$ is less than the escape speed from the core
surface. A simple reason for this is that when the relative speed of bodies 
is less than their surface escape speed 
$l_{min}$ is larger than the impact parameter leading to 
collisions between the cores, so that the inclination strongly 
increases before 
cores have had a significant chance to collide. Thus, all of the late core 
agglomeration resulting in present day terrestrial planets 
occurs only after the disk has become geometrically rather thick, i.e. 
already when $\sigma_i\sim\sigma_e$. 

A similar picture of velocity evolution -- relatively rapid growth of 
inclination compared to the growth of eccentricity --- is expected 
also in the rather general situation of a planetesimal disk 
transitioning from the shear-dominated to the dispersion-dominated 
dynamical regime as a result of gravitational scattering. Indeed, in 
the shear-dominated case eccentricity growth is expected to be much 
faster than the growth of inclination because of the geometry of 
gravitational scattering of planetesimals in a dynamically cold disk. 
Thus, when $\tilde \sigma_e$ becomes comparable to 
unity and continues to grow the condition (\ref{eq:condit}) is  
fulfilled, meaning that very soon after leaving the shear-dominated 
regime the planetesimal disk should rapidly increase its inclination in 
accordance with equations (\ref{eq:sig_e_sol})-(\ref{eq:sig_i_sol_1}). 
This qualitative picture has indeed been observed in calculations of 
planetesimal velocity evolution based on direct N-body simulations 
(Ida \& Makino 1992).

%%%%%%%%%%%%%%%%%%%%%%%%%%%%%%%%%%%%%%%%%%%%%%%%%%%%%%%%%%%
\section{Discussion.}
\label{sect:disc}

In this paper we have explored for the first time a rather special 
dynamical regime of planetesimal velocity evolution represented by 
the condition (\ref{eq:condit}). Previous work towards understanding
planetesimal dynamics has been primarily focused on thick planetesimal 
disks with $i\sim e$ (e.g. Stewart \& Ida 2000; Ohtsuki \etal 2002). 
Thin disks have been considered by Palmer \etal (1993) but their study 
assumed a razor-thin disk, i.e. $i=0$, which precluded them from 
studying a very important aspect of the problem -- excitation of
inclination in a thin disk. They have explored horizontal velocity
excitation in a 2D disk, however, their results cannot be directly 
compared with ours: Palmer \etal computed quantities like
$d v_r^2/dt$ -- growth rate of the radial velocity dispersion of
planetesimals --- which cannot be directly related to our $d\sigma_e^2/dt$
since for the latter one also needs to know the growth rate of 
azimuthal velocity dispersion. 
Nevertheless, their $d v_r^2/dt$ scales linearly with $\sigma_e$, 
in agreement with our equations (\ref{eq:hatH_1}), (\ref{eq:hatH_2}) 
and (\ref{eq:H_1}), (\ref{eq:H_2}) for eccentricity-based scattering 
coefficients. Ida (1990) has also recovered a linear dependence of 
2D horizontal excitation rates on $\sigma_e$ numerically. 
 
The transition between the thin-disk and thick-disk regimes 
of scattering which 
occurs at $\tilde i\sim \tilde i_{crit}$ has not been previously investigated. 
It is known from the studies of thick planetesimal disks that 
scattering coefficients tend to diverge as $\tilde i\to 0$ 
(e.g. Stewart \& Ida 2000). On the other 
hand, Ida (1990) has found numerically that in a purely 2D disk 
horizontal scattering coefficients are finite, which led him to 
a conjecture that these coefficients should 
change discontinuously at $\tilde i=0$. Our present study shows this not 
to be the case. Instead, 3D rates increase with decreasing $\tilde i$
until $\tilde i$ reaches $\tilde i_{crit}$ at which point the 
geometry of scattering 
changes and scattering coefficients smoothly transition to their
2D values. This process is best illustrated in Figure \ref{fig:hat_i} 
where we plot both our thin disk results and the scaling of scattering 
coefficients with $\tilde i$ in the 3D regime. Using analytical expressions 
for various scattering coefficients in Stewart \& Ida (2000) we have 
verified that the magnitudes of 3D scattering coefficients coincide
(up to numerical constant of order unity) with values of our 2D 
scattering coefficients at $\tilde i\sim \tilde i_{crit}$.

Our final comment concerns multiple scattering encounters resulting in
temporary captures such as the
orbit shown in Figure \ref{fig:orb1}. A long time spent by one body in the
vicinity of the Hill sphere of another opens up the possibility of capturing
this body into a distant satellite orbit if some weak additional 
perturbation (e.g. gas drag, collision with/or gravitational perturbation by
an additional passing planetesimal) affects the mutual orbit of the bodies. 
By distant satellite we imply a satellite with separation larger than 
the mutual Hill sphere of the two bodies, and it seems plausible that the 
formation of such a configuration should somehow involve a high velocity
encounter between the two objects (relative speed of a distant satellite 
and its parent body exceeds Hill velocity $\Omega R_H$).

Multiple scattering encounters between low-velocity planetesimals  
potentially leading to the formation of satellites with separations 
less than $R_H$ have been studied by e.g. Iwasaki \& Ohtsuki (2007) 
and Schlichting \& Sari (2008). The high-velocity regime of 
multiple scattering and temporary capture has not yet been explored 
theoretically and is likely to differ from the low-velocity regime. 
In particular, Schlichting \& Sari (2008) found that the probability 
of forming temporary satellite system drops exponentially with the 
duration of the temporary capture in the low-velocity regime, and that 
essentially no temporary capture systems should exist for more than several 
tens of $\Omega^{-1}$. However, in the high-velocity case 
we are finding orbits like the one displayed in Figure \ref{fig:orb1}, which
exhibit temporary capture for more than $10^3\Omega^{-1}$.

Distant satellites have not yet been discovered in planetary systems 
but their dynamics were investigated theoretically by a number of 
authors (Jackson 1913; Lidov \& Vashkov'yak 1994a,b). 
In particular, recently Shen \& Tremaine (2008) 
have demonstrated using a mapping approach that distant satellites 
around some planets (Jupiter, Uranus, and Neptune) are stable on
time scales comparable to the life
time of the Solar System. Temporary captures resulting from 
multiple scattering of dispersion-dominated planetesimals described 
in \S \ref{ssect:single_scat} present one of the possible ways in 
which such distant 
satellites may be formed. The dependence of the efficiency of this 
formation channel on the dynamical state of the planetesimal disk may 
provide us with an important probe of the dynamical characteristics of the 
early Solar System. Needless to say, issues like planetary migration
or possible chaotic epochs of the dynamical evolution of the Solar System
planets (Tsiganis \etal 2005; Gomes \etal 2005) must significantly 
complicate the interpretation of 
future detection (or non-detection) of distant satellites. Nevertheless,
the investigation of their formation efficiency in 
temporary capture events like the one shown in Figure \ref{fig:orb1} 
is a worthwhile exercise.

%%%%%%%%%%%%%%%%%%%%%%%%%%%%%%%%%%%%%%%%%%%%%%%%%%%%%%%%%%%
\section{Summary.}
\label{sect:sum}

We investigated the dynamical evolution of vertically thin, 
dispersion-dominated planetesimal disks with eccentricities and 
inclinations obeying the constraint (\ref{eq:condit}). In this
regime of orbital parameters planetesimals see an 
anisotropic flux of incoming bodies (unlike in the case of thick disks),
which dramatically changes the character of gravitational scattering.
In particular, planetesimal velocity evolution is dominated by 
large-angle scattering events, unlike in the thick disk case. 
We derived analytical expressions for the scattering coefficients
in the thin disk regime and compared them with numerical integrations
of test orbits in the Hill approximation. We found good agreement between 
the two approaches for the eccentricity scattering coefficients, while 
the numerical inclination scattering coefficients significantly differ 
from their analytical analogs. We demonstrated that this discrepancy
is caused by the important role of multiple scattering events not 
captured in our analytical calculations, and by the distant interactions
of planetesimals in their approach phase before close encounter.
Based on these results we have studied the velocity evolution of a population 
of protoplanetary cores in the end of the oligarchic phase and shown that
the initially small inclination of this population grows very rapidly
(exponentially) on a very short timescale. The results of this work 
are useful for understanding the velocity 
evolution of shear-dominated planetesimal disks at the transition to the
dispersion-dominated regime and for the formation of distant satellites 
of planets.

\acknowledgements

We are grateful to Hilke Schlichting for careful reading of the manuscript
and useful discussions. Financial support of this work was provided by 
the Sloan Foundation and NASA grant NNX08AH87G. ZS thanks the Advisory 
Council of the Department of Astrophysical Sciences, Princeton University, 
for generous support through a summer student fellowship. RR acknowledges 
the hospitality of Lebedev Physical Institute and Institute of Space Research
during the completion stages of this work.

\appendix

\section{Scattering coefficients in the 2D regime.}\label{app1}

To compute and analyze scattering coefficients characteristic for thin 
planetesimal disks we utilize an approach developed in Nakazawa \etal 
(1989), Ida \etal (1993), Tanaka \& Ida (1996). For given 
$\tilde h$, $\tilde e$, and $\tilde i$ there are two values of the 
horizontal phase $\tau_{c,\pm}$ and time $t_{c,\pm}$
\ba
\tau_{c}^\pm=\pm\left[\frac{4}{3}\left(\frac{\tilde e^2}
{\tilde h^2}-1\right)^{1/2}-
\Big|\arccos\left(\tilde h/\tilde e \right)\Big|\right],~~~
t_{c}^\pm=\pm\frac{4}{3}\left(\frac{\tilde e^2}
{\tilde h^2}-1\right)^{1/2},
\label{eq:tau+t_pm}
\ea
corresponding to the relative orbit passing through the origin 
(i.e. $x=y=0$) in the zero-inclination case (i.e. $z$ identically 
equal to zero). In the case of 
$\tilde i\sim\tilde e$, passage through the origin also implies that
$z=0$ resulting in a constraint on $\omega$. However, in the case of a
very thin disk with very small but non-zero inclination {\it all} 
values of $\omega$ correspond to roughly the same separation from the 
origin, which is mainly determined by the 
value of $\tau$. Orbits passing close to the origin can be expanded 
about $\tau_c^\pm$ in terms of $\eta^\pm=\tau-\tau_{c}^{\pm}$ with the result 
that prior to interaction one planetesimal approaches another with 
velocity (scaled in Hill units by $\Omega R_H$) 
$\tilde {\bf v}_0=(\tilde v_{0,x},\tilde v_{0,y},
\tilde v_{0,z})$ given by
\ba
\tilde v_{0,x}^{\pm}=\pm\left(\tilde e^2-\tilde h^2\right)^{1/2},~~~
\tilde v_{0,y}^{\pm}=\frac{1}{2}\tilde h,~~~
\tilde v_{0,z}^{\pm}=\tilde i\cos(t_{c}^{\pm}-\omega),
\label{eq:velss}
\ea
and moving on a straight line orbit with the following 
coordinates of the point of closest approach:
\ba
\tilde x_{c}^{\pm}=\pm\left(\tilde e^2-\tilde h^2\right)^{1/2}
\left(\frac{\tilde e^2}{\tilde v_0^2}-1\right)\eta^\pm,~~~
\tilde y_{c}^{\pm}=\frac{1}{2}\tilde h
\left(\frac{\tilde e^2}{\tilde v_0^2}-4\right)\eta^\pm,~~~
\tilde z_{c}^{\pm}=\tilde i\sin(t_{c}^{\pm}-\omega),
\label{eq:coords}
\ea
where $\tilde v_0=|\tilde {\bf v}_0|=
[\tilde e^2-(3/4)\tilde h^2]^{1/2}$. Impact parameter of the 
approach trajectory scaled by $R_H$ is
\ba
\tilde l^\pm=\frac{l^\pm}{R_H}=\left[(\tilde x_{c}^{\pm})^2+
(\tilde y_{c}^{\pm})^2+(\tilde z_{c}^{\pm})^2\right]^{1/2}=
\frac{3}{2}\frac{\tilde h}{\tilde v_0}
\left(\tilde e^2-\tilde h^2\right)^{1/2}\eta_\pm.
\ea
In all these expressions we have neglected terms higher order in $\tilde i$. 

In the two-body approximation that we adopt here gravitational 
interaction of planetesimals changes the straight 
line trajectory into a hyperbola defined as
\ba
r=\frac{l\cos\theta}{\sin\theta+\cos f},~~~\tan\theta=\frac{1}{l v_0^2},
\label{eq:r_hyp}
\ea
where $2\theta$ is the bending angle of the trajectory (angle between the 
incoming and outgoing asymptotes of the orbit) and $f$ is the true 
anomaly of the orbit (angle between the line of focii and a particular 
point on a hyperbola), varying from $\pi/2+\theta$ (incoming) to
$-\pi/2-\theta$ (outgoing). It is trivial to show that in $(x,y,z)\equiv
(x_1,x_2,x_3)$ coordinates this hyperbola can be represented as
\ba
x_i=r\left[\frac{x_{c,i}}{l}\cos (f-\theta)-
\frac{v_{0,i}}{v_0}\sin (f-\theta)\right],~~~i=1,2,3.
\label{eq:x_i}
\ea
Also, conservation of angular momentum allows one to relate 
$f$ and $t$ via
\ba
\frac{df}{dt}=\frac{l v_0}{r^2}.
\label{eq:ang_mom}
\ea

To compute the changes of orbital elements from equations 
(\ref{eq:hdot})-(\ref{eq:iydot}) we also adopt approximation of 
``instantaneous interaction''  meaning that we keep time $t$ fixed
(and equal to $t_{c,\pm}$) throughout the scattering process. This 
approximation works well in high velocity encounters like the ones 
we are considering here because the interaction time is short. 
It allows us to integrate equation 
(\ref{eq:ixdot}) as follows (and all others in analogous fashion):
\ba
\Delta \tilde i_1\approx -\cos t_c\int\limits_{-\infty}^{\infty}
\frac{\partial \phi}{\partial \tilde z}dt=-\frac{\cos t_c}{l\tilde v_0}
\int\limits_{-\pi/2-\theta}^{\pi/2+\theta}r^2
\frac{\partial \phi}{\partial \tilde z}df,
\label{eq:demonstr}
\ea
where we choose a value of $\tau_{c}^{\pm}$ closest\footnote{Ambiguity
in the choice of the origin of our $\tau$-expansion arises when 
$|\tau-\tau_c^+|\sim |\tau-\tau_c^-|$. However, trajectories corresponding 
to these values of $\tau$ do not produce noticeable contribution to the 
scattering coefficients.} to a given value of
$\tau$ and then select a value of $t_c$ corresponding to $\tau_{c}$, 
see equation (\ref{eq:tau+t_pm}). Then, using equations 
(\ref{eq:tau+t_pm})-(\ref{eq:demonstr}) one finds that
\ba
&& \Delta \tilde e_x\approx -\sum\limits_{n=\pm}(\sin t_c^n g_{1}^n+
2\cos t_c^n g_{2}^n),~~~~~
\Delta \tilde e_y\approx \sum\limits_{n=\pm}(\cos t_c^n g_{1}^n-
2\sin t_c^n g_{2}^n),
\label{eq:eres}\\ 
&& \Delta \tilde i_x\approx -\sum\limits_{n=\pm}\cos t_c^n g_{3}^n,
~~~~~\Delta \tilde i_y\approx -\sum\limits_{n=\pm}\sin t_c^n g_{3}^n,
\label{eq:ires}\\ 
&& \Delta \tilde h\approx -2\sum\limits_{n=\pm}g_{2}^n,
~~~~~g_{i}^n=2\frac{\tilde v_i^n+\tilde x_i^n\tilde v_0^3}
{1+(\tilde l^n \tilde v_0^2)^2},
\label{eq:hres}
\ea
where $i=1,2,3$ stands for $x,y,z$, correspondingly.

In Figure \ref{fig:pars_of_tau} we compare analytical and numerical results 
for the changes of various orbital elements in the thin-disk limit $\tilde 
i\ll 1\ll \tilde e$, which is of interest for us here. 
As has been previously shown by Tanaka \& Ida (1996), analytical results
match numerical ones for most values of $\tau$ if one shifts the 
origin of $\tau$ by a small amount
$d\tau\ll 1$. This shift arises from the distant interaction 
of the two planetesimals before they have experienced close encounter.
Such a shift in $\tau$ does not affect in any way our calculation of
scattering coefficients averaged over $\tau$. 

A striking feature of Figure \ref{fig:pars_of_tau} is the existence of narrow 
intervals of $\tau$ in which numerical results strongly deviate in
seemingly chaotic fashion from the analytical ones. Deviations of 
$\Delta \tilde e$, can be of order $\Delta \tilde e$ 
itself, but the discrepancy between the numerical and analytical 
$\Delta \tilde i$ exceeds analytical value of $\Delta \tilde i$
by orders of magnitude for some values of $\tau$. The implications 
of these deviations are discussed in more detail in \S 
\ref{sect:numerics}.

Neglecting for now this additional complication we average 
expressions (\ref{eq:eres})-(\ref{eq:ires}) over $\omega$ and $\tau$
and finally arrive at equations (\ref{eq:de2_omtau})-(\ref{eq:idi_omtau}).

\section{Role of distant encounters.}\label{app2}

Let $\omega_0$ be the initial relative vertical phase of two
planetesimals at infinity and $\omega$ be the value of this phase 
right before the close encounter. Their difference 
$\delta\omega_{dist}=\omega-\omega_0$ is small but nonzero because of the
distant interaction of planetesimals preceding their close encounter.
Previously we assumed $\omega$ to be equal to $\omega_0$ (i.e. 
$\delta\omega_{dist}=0$) thus neglecting the effect of distant interaction. 
Let us now see how the fact that $\delta\omega_{dist}\neq 0$ affects 
calculation of $\langle\tilde {\bf i}\cdot\Delta\tilde 
{\bf i}\rangle_{\omega}$. 

Using equations (\ref{eq:velss}), (\ref{eq:coords}), (\ref{eq:ires}), 
(\ref{eq:hres}) we find
\ba
\tilde {\bf i}\cdot\Delta\tilde {\bf i}=\tilde i\cos\omega_0\Delta \tilde i_x+
\tilde i\sin\omega_0\Delta \tilde i_y=-2\tilde i^2\sum\limits_{n=\pm}
\frac{\cos(t^n_c-\omega)\cos(t^n_c-\omega_0)+\tilde v_0^3
\sin(t^n_c-\omega)\cos(t^n_c-\omega_0)}{1+(l^n \tilde v_0^2)^2}.
\label{eq:idi1}
\ea
If we now average this expression over $\omega_0$ we find (recall that 
$l^\pm$ and $\tilde v_0$ are virtually independent of $\omega_0$ in the
thin-disk regime  when $\tilde i\ll \tilde e$)
\ba
\langle\tilde {\bf i}\cdot\Delta\tilde {\bf i}\rangle_\omega\approx
-\tilde i^2\sum\limits_{n=\pm}\frac{1-2\tilde v_0^3\hat{\delta\omega_{dist}}}
{1+(l^n \tilde v_0^2)^2},
\label{eq:idi_om_av}
\ea
where 
\ba
\hat{\delta\omega_{dist}}=\frac{1}{2\pi}\int\limits_0^{2\pi}
\delta\omega_{dist}(\omega_0)\cos^2(t^n_c-\omega_0)d\omega_0.
\label{eq:domega}
\ea
If distant interaction prior to encounter were not taken into account
then $\delta\omega_{dist}=0$ and equation (\ref{eq:idi_om_av}) would be missing
the second term in the numerator because averaging over $\omega_0$ would kill 
this term completely. However, when distant interaction and the possibility 
of non-zero $\delta\omega_{dist}$ are allowed for the omission of the second 
term may not be justified even if $|\delta\omega_{dist}|\ll 1$ because 
$\tilde v_0^3\gg 1$ in the situation that we consider. Then it may 
be possible that the product $\tilde v_0^3\hat{\delta\omega_{dist}}\gtrsim 1$
and dominates the numerator of equation (\ref{eq:idi_om_av}), which makes 
our neglect of distant interaction in calculation of
$\langle\tilde {\bf i}\cdot\Delta\tilde {\bf i}\rangle_\omega$ 
unjustified. Thus, distant interaction of planetesimals can indeed
explain the discrepancy between analytical and numerical values of 
$\langle\tilde {\bf i}\cdot\Delta\tilde {\bf i}\rangle_\omega$ observed 
in Figure \ref{fig:omega_av}.

\end{document}